\documentclass[manuscript,screen]{acmart}
\usepackage{amsmath}
\usepackage{framed}
\usepackage{multirow}
\usepackage{placeins}
\usepackage{comment}
\usepackage{listings}

\renewcommand\footnotetextcopyrightpermission[1]{}

\AtBeginDocument{%
  }

\usepackage{xcolor} 

\begin{document}


\title{Small Language Models Can Use Nuanced Reasoning For Health Science Research Classification: A Microbial-Oncogenesis Case Study}



\author{Muhammed Muaaz Dawood}
\authornote{Both authors contributed equally to this research.}
\authornote{Also with Infectious Diseases and Oncology Research Institute (IDORI), Faculty of Health Sciences, University of the Witwatersrand}
\email{2425639@students.wits.ac.za}
\orcid{0009-0007-0408-5661}
\affiliation{%
  \institution{School of Computer Science and Applied Mathematics, University of the Witwatersrand}
  \country{Johannesburg, South Africa}
}


\author{Mohammad Zaid Moonsamy}
\authornotemark[1]
\authornotemark[2]
\authornote{Corresponding author}
\affiliation{%
  \institution{School of Computer Science and Applied Mathematics, University of the Witwatersrand}
  \country{Johannesburg, South Africa}}

\email{2433079@students.wits.ac.za}
\orcid{0009-0006-5321-8179}

\author{Kaela Kokkas}
\authornotemark[2]
\orcid{0009-0002-9916-0049}
 \affiliation{%
   \institution{Department of Clinical Microbiology and Infectious Diseases, University of the Witwatersrand}
   \country{Johannesburg, South Africa}
 }

 \author{Hairong Wang}
 \affiliation{%
   \institution{School of Computer Science and Applied Mathematics, University of the Witwatersrand}
   \country{Johannesburg, South Africa}
 }
 \orcid{0000-0001-8770-5916}

\author{Robert F. Breiman}
 \affiliation{%
   \institution{Infectious Diseases and Oncology Research Institute (IDORI), Faculty of Health Sciences, University of the Witwatersrand}
   \country{Johannesburg, South Africa}
 }
 \orcid{0000-0002-7099-2936}

\author{Richard Klein}
 \affiliation{%
   \institution{School of Computer Science and Applied Mathematics, University of the Witwatersrand}
   \country{Johannesburg, South Africa}
 }
 \orcid{0000-0003-0783-2072}
 
 \author{Emmanuel K. Sekyi}
 \affiliation{%
   \institution{OncoVectra}
   \country{London, United Kingdom}
 }
 \orcid{0000-0002-8231-6935}

 \author{Bruce A. Bassett}
 \authornotemark[2]
 \affiliation{%
   \institution{Wits MIND Institute and School of Computer Science and Applied Mathematics,  University of the Witwatersrand}
\country{Johannesburg, South Africa}}


\orcid{0000-0001-7700-1069}





\renewcommand{\shortauthors}{Dawood et al.}


\begin{abstract}
 
Artificially intelligent (AI) co-scientists must be able to sift through research literature cost-efficiently while applying nuanced scientific reasoning. We evaluate Small Language Models (SLMs, $\leq$ 8B parameters) for classifying medical research papers. Using literature on the oncogenic potential of HMTV/MMTV-like viruses in breast cancer as a case study, we assess model performance with both zero-shot and in-context learning (ICL; few-shot prompting) strategies against frontier proprietary Large Language Models (LLMs). Llama 3 and Qwen2.5 outperform GPT-5 (API, low/high effort), Gemini 3 Pro Preview, and Meerkat in zero-shot settings, though trailing Gemini 2.5 Pro. ICL leads to improved performance on a case-by-case basis, allowing Llama 3 and Qwen2.5 to match Gemini 2.5 Pro in binary classification. Systematic lexical-ablation experiments show that SLM decisions are often grounded in valid scientific cues but can be influenced by spurious textual artifacts, underscoring need for interpretability in high-stakes pipelines. Our results reveal both promise and limitations of modern SLMs for scientific triage; pairing SLMs with simple but principled prompting strategies can approach performance of the strongest LLMs for targeted literature filtering in co-scientist pipelines.

\end{abstract}





\maketitle

\section{Introduction}

Artificial Intelligence (AI) co-scientists represent a promising new paradigm in scientific research that could revolutionize how researchers search, process and ideate within the exponentially growing research literature~\cite{bornmann2021growth}, with an estimate of more than  2 million new papers added to the corpus of human knowledge each year~\cite{hanson2024strain}. In addressing a generic question that requires deep understanding of the existing research, an AI co-scientist must undertake three broad steps: (1) {\bf Discover:} scan the entire literature for relevant research, (2) {\bf Filter:} narrow this initial set of results down to a highly relevant subset, and (3) {\bf Understand:} process this final subset in as much depth as the AI can manage.

Current commercial approaches, such as the {\em Deep Research} modes of Gemini and GPT-5~\cite{google_gemini_deep_research, openai_chatgpt5_deep_research}, can execute these tasks but function as opaque, resource-intensive black boxes. Their tool calls, search strategies, and trade-offs between cost and performance are hidden from users, limiting transparency, reproducibility, and user control in scientific workflows. For many research questions, relying solely on large language models (LLMs) for all three stages can require substantial computational resources, making them impractical for filtering thousands of research papers while also raising concerns about energy consumption and environmental sustainability.

We argue that Small Language Models (SLMs, which we define here as models with $\leq$ 8B parameters) offer a cost-efficient and transparent solution for step (2): filtering. Within a co-scientist pipeline, SLMs can efficiently reduce large search results to a core subset of highly relevant studies, which can then be handed off to frontier LLMs for deeper semantic analysis. The success of this approach depends critically on whether SLMs can accurately classify research papers—often using only titles and abstracts—into relevant and irrelevant categories. Achieving this requires reasoning over nuanced inclusion criteria that go well beyond traditional NLP (Natural Language Processing) classification tasks such as sentiment analysis or relevance detection based on simple embedding features~\cite{mikolov2013efficient, pennington2014glove}. The central question, therefore, is whether SLMs can perform such nuanced classification reliably enough to be useful in scientific research pipelines.

To explore this, we focus on the field of microbial oncogenesis, in which microbes such as viruses, bacteria, and fungi induce or exacerbate cancer development. New discovery of microbial facilitation of one or more cancers (especially those linked to high mortality and disability) has immense life- and cost-savings potentials, via developing (or using existing) affordable tools and strategies to prevent the infection. This domain presents a challenging test for SLMs, as the classification criteria are highly nuanced and the evidence spans multiple disciplines, including oncology, microbiology, genetics, pathology, epidemiology, and immunology. The difficulty of efficiently applying limited available resources for laboratory and epidemiologic research towards discovery of identifying novel and high burden meaningful microbe-cancer pairs (MCPs) is compounded by the overwhelming volume of biomedical literature, which doubles approximately every nine years~\cite{bornmann2015growth}. Manual review is increasingly impractical, while failure to systematically examine available literature risks overlooking critical evidence. Conventional keyword-based search engines, though sensitive, yield numerous false positives that require extensive manual curation~\cite{Zhou2022}. Efficient identification of relevant studies therefore requires automated screening methods capable of expert-level reasoning.

We focus on the debated association between the \textbf{MMTV-like virus and breast cancer}, which presents a rich and nuanced classification challenge representative of real-world biomedical literature screening tasks. Using this case as a specific but generalizable example, we systematically evaluate the ability of open-weight SLMs to classify the relevance of medical research papers. More broadly, this work aims to establish a foundation for scalable, automated discovery of high-priority microbial–cancer associations across the biomedical literature. To guide this evaluation, we investigate four primary research questions:

\begin{enumerate}
    \item Can small language models ($\leq$ 8B parameters) perform accurate and nuanced classification of medical research papers based on complex reasoning criteria?
    \item How do different prompting strategies—specifically zero-shot, various few-shot methods, and advanced prompt optimization—affect SLM classification performance?
    \item How does the performance of these small, locally-runnable models compare to that of a state-of-the-art frontier models (e.g. Gemini 2.5 Pro and GPT-5) serving as zero-shot baselines?
    \item What lexical features drive the classification choices of the SLMs and are these grounded in appropriate causal medical reasoning or not? 
\end{enumerate}

To support this evaluation, we curated and publicly release a dataset of 100 medical research articles (title-abstract pairs) with expert-validated labels indicating relevance to MMTV-like viruses as a potential cause of breast cancer.

In the remainder of this paper, Sections~\ref{sec:background} and~\ref{sec:related} establish the context of microbial oncogenesis and position our work within the landscape of automated literature screening. Section~\ref{sec:method} outlines our experimental framework, including expert-validated dataset construction and in-context learning strategies. Section~\ref{sec:results} reports classification performance across SLMs and frontier LLMs, analyzes the impact of prompting strategies, and investigates model reasoning through perturbation analysis. Section~\ref{sec:limitations} discusses study limitations. Section~\ref{sec:conclusion} concludes the paper, and Section~\ref{sec:futurework} outlines directions for future research.

\section{Background}\label{sec:background}

\subsection{Microbe-Cancer Associations}
The development of vaccines for HPV and hepatitis B in the 1980s, which now prevent an estimated 800,000 annual deaths from cervical and liver cancers, established a powerful new paradigm for cancer prevention~\cite{Watts2023, WorldHealthOrganisationWHO2022}. This success highlighted the potential of identifying and targeting causal Microbe-Cancer Pairs (MCPs) to develop further interventions~\cite{ToumivanDorsten2025}. Despite this promise, the discovery of new oncogenic microbes has been slow, with only nine confirmed in the six decades following the first~\cite{Rowe2014, IARC2012}. This is partly due to the vast and sometimes conflicting body of scientific literature that must be assessed to establish causal links.

One such proposed causal MCP is MMTV-like virus and breast cancer. MMTV-like virus was named for its genetic similarity to Mouse Mammary Tumor Virus (MMTV), first identified in mouse mammary tissue and established as the main cause of mammary cancer in mice. MMTV has long been used to induce mammary tumors in mice, which closely resemble certain human breast tumors~\cite{Ross2010, Lawson2019}. These parallels prompted questions about whether MMTV—or a related virus—might infect humans and contribute to breast cancer. A virus nearly identical to MMTV was subsequently discovered in human breast cancer tissue, termed MMTV-like virus and later renamed Human Mammary Tumor Virus (HMTV) due to its association with human mammary tumors~\cite{Axel1972}. However, decades of research across multiple disciplines have produced conflicting evidence about HMTV/MMTV-like virus’s etiological role in human breast cancer, making comprehensive synthesis of the literature increasingly difficult.

\subsection{Automated Literature Screening in Biomedical Sciences}
The difficulty of manually synthesizing evidence for cases like MMTV-like virus reflects a broader challenge. The exponential growth of biomedical publications has driven the development of automated tools to support systematic reviews and evidence synthesis. Early approaches relied on rule‑based and statistical classifiers—such as naïve Bayes or support vector machines—trained on manually curated keywords and abstracts to filter large corpora ~\cite{tsafnat2014systematic}. However, keyword‐based methods frequently suffer from low precision—due to polysemy, homonyms, and broad term matches—despite often achieving high sensitivity by retrieving a wide range of potentially relevant articles ~\cite{bramer_2016_comparing}. This results in the retrieval of many irrelevant papers, necessitating substantial manual screening efforts to identify truly relevant studies.

Embedding-based retrieval techniques leverage transformer encoders like SciBERT and BioBERT to map text into dense semantic vectors, then perform nearest-neighbor search in that space ~\cite{beltagy2019scibert,lee2020biobert}. These methods substantially improve both precision and sensitivity over classical approaches by capturing contextualized semantics. However, because they rely solely on vector similarity, they struggle when relevance is defined by nuanced inclusion criteria that go beyond surface-level semantic similarity. This motivates the use of language models, which can reason over entire abstracts and apply such criteria more explicitly, potentially enabling more reliable screening decisions.

\subsection{Language Models for Screening and Classification}
Large Language Models (LLMs) have recently shown promise in automating the screening phase of systematic reviews~\cite{huotala2024promise}. However, most studies to date rely on very large or proprietary models, with the former demanding substantial computational resources and the latter requiring paid API access~\cite{huotala2024promise,delgadochaves2025transforming}. Proprietary models additionally offer limited transparency into their operations and updates, making it difficult to ensure reproducibility across studies. These resource and transparency constraints limit accessibility for many research groups, driving growing interest in smaller, open-weight models that can be deployed locally.

Open‑weight models in the 8 billion‑parameter range (e.g., Llama (8B)~\cite{llama3modelcard}, Qwen2.5 (7B)~\cite{qwen2.5}) strike a balance between inference efficiency and representational power, enabling on‑premise deployment~\cite{touvron2023llama,zhao2024qwen}. Recent surveys indicate that such relatively small models can achieve competitive results in specialized settings and highlight ongoing challenges in trustworthiness and adaptation to domain‑specific tasks~\cite{wang_comprehensive_2024}. While large‑scale benchmarks have established their raw capabilities, systematic evaluations on specialized biomedical screening tasks remain limited.

\subsection{In-Context Learning}
A simple, yet powerful, first step for adapting these language models to biomedical screening tasks is In-Context Learning (ICL)~\cite{brown2020language}. ICL enables language models to perform tasks by conditioning on examples provided directly within the prompt (few-shot examples), without modifying the model's parameters or requiring retraining. This approach encompasses zero-shot learning, where only task instructions are provided; one-shot learning, which includes a single demonstration; and few-shot learning, where multiple examples guide the model's behavior. Few-shot learning has been shown to substantially improve LLM performance across diverse tasks~\cite{brown2020language}, including biomedical applications~\cite{fung2025few}. However, this performance is highly sensitive to which examples are selected and how they are ordered. Poor demonstration choices can lead to significant accuracy degradation, yet manual curation becomes infeasible as the search space grows~\cite{khattab2024dspy}. This challenge has motivated the development of automated few-shot selection strategies and optimization frameworks such as AdalFlow~\cite{adalflow} and DSPy~\cite{khattab2024dspy}.

\section{Related Work}\label{sec:related}

Several recent studies have investigated the application of large language models (LLMs) to literature screening, particularly within biomedical domains. Dennstädt et al.~\cite{dennstadt2024} explored the use of openly available LLMs with structured prompts and Likert-scale scoring for biomedical abstract classification. While their study demonstrates that such models can flexibly support relevance assessments, they also report substantial trade-offs between sensitivity and specificity and note that performance is highly dependent on prompt design and dataset characteristics. Their use of a multi-point Likert scale differs from our setting, which directly classifies abstracts into three categories (\emph{relevant}, \emph{somewhat relevant}, \emph{irrelevant}), offering a more interpretable and practical decision scheme. Although their work included small open-weight models and biomedical data, it did not examine in-context learning (ICL) as a strategy for improving classification.

Delgado-Chaves et al.~\cite{delgadochaves2025transforming} conducted a broad evaluation of 18 LLMs across multiple screening datasets. Their findings show that LLMs can reduce reviewer workload by up to 93\% and that model size alone is not a reliable predictor of performance. They further highlight the importance of clearly defined inclusion and exclusion criteria, and investigate hybrid strategies in which LLM predictions are combined with downstream classifiers. Their work also explores retrieval-augmented generation (RAG) for improving screening efficiency. While they included some smaller models in their comparisons, the majority of their evaluation focused on large proprietary or API-based systems, and they did not assess the use of ICL for screening.

Huotala et al.~\cite{huotala2024promise} examined both partial and full automation strategies for abstract screening. While text simplification via GPT-3.5 reduced reviewer time without improving accuracy, they found that GPT-4 could achieve near-human screening performance when combined with one-shot, few-shot, or chain-of-thought prompting. Their study emphasizes the value of ICL for improving screening accuracy, but their evaluation focuses exclusively on large proprietary models, leaving open the question of how smaller open-weight models perform under similar prompting strategies.

While Huotala et al. \cite{huotala2024promise} explored in-context learning (ICL) for systematic review screening, they manually selected seed papers for few-shot prompting. To automate this labor-intensive selection process, programmatic optimization frameworks like DSPy~\cite{khattab2024dspy} have emerged. Khattab et al~\cite{khattab2024dspy} demonstrated that DSPy-compiled pipelines enable smaller, open-source models (like Llama 2-13B) to substantially outperform standard few-shot prompting and achieve performance competitive with large proprietary models using expert-written prompts. This emphasis on optimizing bootstrapped demonstrations as a key to performance \cite{opsahl2024optimizing} has been adopted in subsequent work. For instance, studies by Sarmah et al.~\cite{sarmah2024comparative} and Lemos et al.~\cite{lemos2025time} utilized DSPy's Bootstrap Few-Shot with Random Search (BFS-RS) optimizer with large models (GPT-4o and Llama 3.1-70B) to achieve improved results on their respective tasks. However, this powerful programmatic approach has not yet been applied to assess its value in the context of small, open-weight models for specialized domains like biomedical screening.

Task-specific adaptations have also been explored. A recent question-answering (QA) framework reformulated screening as a series of structured yes/no questions about inclusion criteria~\cite{qa_framework_paper}. This approach achieved high recall and reduced the number of abstracts requiring manual review, demonstrating the flexibility of LLMs for task reformulation. However, it relied entirely on proprietary models and did not employ ICL-based strategies, limiting its applicability for resource-constrained settings.

In summary, existing studies demonstrate that LLMs can substantially improve efficiency in literature screening, but each investigates only part of the problem space: some focus on biomedical domains, others on small models, and others on ICL and prompting strategies. Our work brings these dimensions together by systematically evaluating \textbf{small, open-weight LLMs} ($\leq$8B parameters) for \textbf{biomedical literature screening} under \textbf{few-shot prompting and prompt optimization}. This combined focus addresses gaps in prior work by testing whether resource-efficient models can deliver competitive performance in biomedical screening tasks, while also assessing the relative effects of manual few-shot learning and programmatic prompt optimization.

\section{Methodology}\label{sec:method}

\subsection{Dataset}

We curated a gold-standard dataset of 100 \cite{Kincaid2018, Ahmad2023, Lee1995, Segev1985, May1986, Shoenfeld1987, Bansal1990, O1993, Pogo2010, Park2011, Melana2010, Indik2007, Goedert2006, Lehrer2019, Jin2002, Jayatilleke2023, Liu2023, Dhokotera2024, Maroga2023, Dewan2025, Ramalingam2025, Alaouna2025, Dix-Peek2023, Prabhu2024, Tang2022, Armstrong2023, Achilonu2022, Bhadra2006, Santiago-Rivera2025, Gong2025, Simpson2003, Otten1988, Glover1989, Matsuzawa1995, Yang2024, Lee2022, Gomes2023, Yao2023, Tang2016, Dreyer2023, Zecca2024, Ishikawa2008, Mu2024, Kim2024, Xu2024, Auricchio2019, Mahanty2003, Abrahamsson2024, Xiao2017, Mazzanti2011, Trejo-Avila2011, Nojiri2025, Gentry2025, Wang2025, Willsey2025, TheATLASCollaboration2024, Kar2025, Chen2025, Pleguezuelos-Manzano2020, Faria2025, Giordano2025, Kim2025, Copland2024, Hashizume2025, Wang_2_2025, Ernst2014, Kanwal2017, vanLeeuwen2024, Arnaudova2024, Callahan2012, Tabriz2013, Kulkarni2013, Morales-Snchez2013, Mazzanti2015, Cedro-Tanda2014, Reza2015, Christenson2017, Naushad2017, Paul2020, Obr2018, Perzova2017, Shariatpanahi2017, AlDossary2018, Kim2019, Naccarato2019, Lessi2020, Pereira2020, Kaul2021, Zhang2021, FawadKhalid2021, Kwa2021, Velazquez2021, Martnez-Nieto2021, Gupta2021, Takeuchi2021, Stewart2022, Gupta2022, Khalid2023, Pannone2014, James2024} medical research papers (titles and abstracts) with expert-adjudicated ground-truth classifications, rather than crowdsourced or LLM-generated labels. We specifically focused on titles and abstracts, as the sheer volume of biomedical literature makes it impractical to process full papers during the initial filtering stage. Titles and abstracts serve as the first-stage screening layer, with full-text analysis reserved for the downstream processing of the filtered subset. Labeling was conducted through detailed, iterative discussions within our team, led by an experienced expert in infectious diseases and oncology (RFB). It is well-recognized that obtaining reliable annotations from medical experts is challenging, even for tasks such as literature screening, as it requires highly specialized domain knowledge. Accurate classification in this domain requires deep, interdisciplinary reasoning spanning virology, oncology, and pathology, making reliable labeling both time-intensive and costly. Generating large quantities of equivalently vetted examples is therefore infeasible. Prioritizing scale over accuracy risks creating noisy benchmarks that obscure true reasoning capabilities. Consequently, we \textbf{prioritized expert reliability and label fidelity}, framing this collection as a \textbf{high-quality evaluation set} designed to rigorously assess model performance. Each paper was labeled as \emph{relevant} (38 papers), \emph{somewhat relevant} (15 papers), or \emph{irrelevant} (47 papers) based on the following question:
\begin{quote}
    How relevant is this article in determining whether it is plausible that HMTV/MMTV-like virus can cause breast cancer in humans?
\end{quote}
Detailed classification criteria and the prompt are provided in Section~\ref{sec:prompts}.

We split the dataset into training, validation, and test sets using stratified sampling with proportions of 20\%, 30\%, and 50\%, respectively. Given the small dataset size and few-shot learning context, we allocated a larger portion to the test set to ensure robust evaluation while retaining sufficient samples for training and validation. Stratification maintained consistent class distributions across all splits. All evaluations, including zero-shot experiments, were conducted on the test set to assess the model's ability to classify unseen papers.

For data extraction, we used GROBID (GeneRation Of BIbliographic Data) \cite{lopez2009grobid} to process PDF files, achieving 80\% extraction accuracy of title and abstracts. Remaining titles and abstracts were manually extracted. The final dataset was formatted as JSON, containing paper titles, abstracts, and associated labels. We publicly release this dataset to support further research in oncology literature classification. 

\subsection{Models}\label{sec:models_times}

We focus on small language models ($\leq$ 8B parameters) due to their cost efficiency, open-weight availability, and fast inference speed. This approach prioritizes scalability, as our AI co-scientist workflow must be able to process thousands of biomedical papers, making high‑latency or memory‑intensive models impractical. We utilized HuggingFace’s Open LLM Leaderboard to shortlist models including Llama 3 (8B)~\cite{llama3modelcard}, Qwen2.5 (7B)~\cite{qwen2.5}, Qwen3 (8B)~\cite{qwen3}, DeepSeek‑R1 (7B)~\cite{deepseekai2025deepseekr1incentivizingreasoningcapability}, and Mistral (7B)~\cite{2023arXiv231006825J}.

To evaluate these open-weight models, we hosted them locally using the Ollama\footnote[1]{https://ollama.com/} API for all experiments. Timing benchmarks showed that Qwen2.5 processed 100 papers in 69.68 seconds (0.70s per paper) and Llama 3 in 76.14 seconds (0.76s per paper), while Qwen3 and DeepSeek‑R1 were significantly slower at 2591.66 seconds (25.92s per paper) and 1420.19 seconds (14.20s per paper), respectively — roughly 37$\times$ and 20$\times$ slower than Qwen2.5. At scale, such differences are critical: processing 10,000 papers would take under 2 hours with Qwen2.5 or Llama 3 but nearly 3 days with Qwen3. These time lags are significantly exacerbated if one wants to employ In-Context Learning (ICL) searches. For example, DSPy’s Bootstrap Few-Shot with Random Search (BFS-RS) optimizer becomes increasingly memory- and time-intensive as the number of candidates grows. We evaluated up to 150 candidates, and due to race conditions and non-deterministic results when using more than one thread, we restricted execution to a single thread to ensure rigorous and reproducible results. Consequently, even for Qwen2.5 and Llama 3, the total analysis time for hyperparameter tuning was between 100 and 200 hours, making similar hyperparameter tunings for DeepSeek-R1 and Qwen3 infeasible.  

While Mistral 7.2B achieved similar runtime performance to Qwen2.5 and Llama 3, we prioritized Llama 3 based on its superior performance on medical reasoning benchmarks~\cite{kim2025small}. As a result, we only selected Llama 3 and Qwen2.5 for detailed study. For benchmarking comparisons, we additionally evaluated several frontier LLMs—Gemini 2.5 Pro~\cite{comanici2025gemini}, Gemini 3 Pro Preview~\cite{google_deepmind_gemini3_2025}, GPT-5 (low- and high-effort reasoning API variants)~\cite{openai_gpt5_2025}, and GPT-5 Thinking-Mini (UI)~\cite{openai_gpt5_mini_CLI_2025}—as well as Meerkat (7B)~\cite{kim2025small}, a model based on Mistral-7B and fine-tuned on medical data~\cite{kim2025small}. All models were run in a zero-shot setting. Model temperatures were fixed at 0, except for GPT-5 (which does not allow manual temperature control) and Gemini 3 Pro Preview (for which a temperature of 1.0 is recommended for optimal performance). Finally, details of the hardware used for the timing benchmarks are provided in Appendix~\ref{app:hardware_timing}.

\subsection{Evaluation Metrics}

We evaluated model performance using standard classification metrics, including precision, sensitivity (recall), specificity, accuracy and $F_{\beta}$ scores (with $\beta = 0.2, 1, 5$). $F_{0.2}$ prioritizes precision while $F_5$ prioritizes recall. We included these to cover two extremes: (i) if a given search returns few relevant papers, missing relevant papers is a serious issue whilst misclassifying irrelevant papers as relevant has minimal cost implications (since the downstream LLM will remove them with high probability). However, (ii) if thousands of papers are returned from the initial search, misclassifying a significant fraction of irrelevant papers as relevant will significantly increase the downstream computational and financial costs while missing some relevant papers is likely to be less serious since the knowledge they contain is more likely to be encoded in other relevant papers.

\subsection{The Classification Prompt}
\label{sec:prompts}
The prompt we used across all  experiments is shown below:

\begin{framed}
\textbf{Classification Prompt:}

How relevant is this article in determining whether it is plausible that HMTV/MMTV-like virus can cause breast cancer in humans?

\medskip

A \textbf{relevant} article would fulfill either of the following criteria:
\begin{itemize}
    \item Gives causal evidence for whether or not HMTV/MMTV-like virus or its species/family of microbes cause breast cancer or other cancers, either in humans or in animal models.
    \item Contains a theoretical model for how HMTV/MMTV-like virus might cause breast cancer, irrespective of whether there is experimental evidence.
\end{itemize}

A \textbf{somewhat relevant} article would fulfill either of the following criteria:
\begin{itemize}
    \item Mentions both HMTV/MMTV-like virus and breast cancer, but does not provide any causal arguments or evidence for the role of the virus in causing breast cancer.
    \item Contains evidence examining whether other microbes cause breast cancer, if the findings can plausibly generalize to HMTV/MMTV-like virus and breast cancer.
\end{itemize}

An \textbf{irrelevant} article would be a paper that does not fulfill the criteria for the relevant or somewhat relevant classifications.

\medskip

\textbf{FORMAT:} In your answer, use only \textsc{Relevant}, \textsc{Somewhat Relevant}, \textsc{Irrelevant} or \textsc{No Classification} as your response.
\end{framed}
This prompt went through several iterations for clarity. While these definitions may cause debate among field experts, their complexity enables thorough assessment of LLMs’ ability to handle nuanced, field-specific prompts. In Appendix \ref{sec:logic} we provide a fully general, formal logic version of this prompt illustrating the complexity and nuance of the prompt predicates and its generality. 

Classifying papers using these criteria requires multiple layers of consideration and application of field-specific knowledge. First, it is necessary to identify whether the paper discusses a microbe, and if so whether the microbe is HMTV/MMTV-like virus or a synonym thereof, a related virus within the same family or species, or an entirely unrelated microbe. Similarly, it is necessary to identify whether the paper discusses a cancer, and if so, whether it is breast cancer or a subtype thereof, or another type of cancer entirely.

Since we want to evaluate causal links, classification requires more than simply identifying which papers mention both HMTV/MMTV-like virus and breast cancer, as papers that do this may only be somewhat relevant by the above criteria because it does not provide causal evidence and/or a theoretical model. Conversely, research that does not mention HMTV/MMTV-like viruses or breast cancer could be relevant if it provides causal evidence that a virus within the same species or family does (or does not) cause a different cancer. Additionally, the relevant classification definition prioritizes inclusion of HMTV/MMTV-like virus and its relatives over breast cancer. If the paper focuses on a Microbe-Cancer Pair (MCP) consisting of HMTV/MMTV-like virus or its relatives and a cancer other than breast cancer, the paper is still considered {\em relevant}. However, if the paper focuses on another, unrelated microbe and breast cancer, it is considered {\em somewhat relevant}. 

Finally, two components of the prompt require additional nuanced interpretation and inference to ensure accurate classification. First, a criterion of the relevant classification calls for “causal evidence” for or against the MCP. This necessitates not only interpreting the paper’s findings but also extrapolating what findings could reasonably support or refute a causal relationship. In particular, evidence against a causal relationship is distinct from no evidence about a causal relationship. 

Second, a criterion of the somewhat relevant classification includes papers involving unrelated microbes and breast cancer, provided their findings can “plausibly generalize” to HMTV/MMTV-like virus. Applying this requires an understanding of both the unrelated microbe, HMTV/MMTV-like virus, and the paper’s results to determine whether the evidence provided could reasonably apply to HMTV/MMTV-like virus and breast cancer.

\subsection{Few-Shot Learning}\label{sec:few-shot-methods}
As a baseline, we evaluated the models' zero-shot performance by providing the prompt from Section~\ref{sec:prompts}, which required classifying papers solely based on their title and abstract. We tested different zero-shot implementations: Vanilla, Rationale (where the model is prompted to produce an explanation alongside the predicted label), AdalFlow~\cite{adalflow} (a framework for building and auto-optimizing LLM task pipelines with support for zero- and few-shot learning), and DSPy~\cite{khattab2024dspy} implementations with Predict and Chain-of-Thought (CoT) modules. The DSPy Predict module enables direct structured input–output inference, while CoT explicitly prompts the model to reason step-by-step before producing the final output. This established a reference point for assessing the effectiveness of few-shot strategies. We then examined few-shot learning approaches that incorporate a small number of title–abstract–label examples into the prompt. The techniques were implemented using AdalFlow, except BFS-RS which was implemented using DSPy. The five few-shot ICL methods used for example selection are described below.

\begin{description}
    \item[Random Few-Shot] Examples were randomly selected without regard for class balance. 
    \item[Fair Few-Shot] This approach ensures balanced class representation by selecting equal numbers of examples from each class. 
    \item[Central Examples Few-Shot] Title--abstract pairs are converted into vector representations using an embedding model \footnote[2]{microsoft/BiomedNLP-PubMedBERT-base-uncased-abstract-fulltext}. For each class, we then computed its centroid —the average semantic position—and selected the nearest example as the class prototype to include in the prompt.  This provides archetypical reference examples for the models from each class, which can improve classification consistency~\cite{huang2019centroid}.
    \item[Closest Examples Few-Shot] Employing the same embedding process, this method selects demonstrations based on their proximity to a new input in the vector space. To mitigate class imbalance, we ensured the selected examples were drawn equally from each class. This approach provides the model with prompt demonstrations that are both semantically similar to the query and representative across classes, supporting balanced and contextually grounded classification~\cite{xu2023knn}.
    \item[Bootstrap Few-shot with Random Search (BFS-RS)] For this strategy, we employed DSPy's bootstrapping capabilities instead of AdalFlow, due to its more intuitive interface and broader selection of optimization strategies. This choice was further motivated by AdalFlow's own reliance on DSPy-inspired bootstrapping techniques \cite{adalflow}. DSPy provides a suite of optimizers to enhance performance on few-shot learning tasks. Among them, Bootstrap Few-Shot (BFS) automates the selection of high-quality examples—known as demonstrations—to include in the prompt. It employs a teacher model (typically identical to the model being optimized—the student) to generate reasoning traces and predictions for training inputs. When the teacher's prediction matches the ground-truth output, the corresponding input–reasoning–output sequence is stored as a \textbf{bootstrapped demonstration}. The optimizer iterates through the training set until it collects the required number of bootstrapped demonstrations.
    
    A limitation of BFS is its sensitivity to the ordering of samples in the training data. To address this, the Bootstrap Few-Shot with Random Search (BFS-RS)~\cite{khattab2024dspy} optimizer extends the base algorithm by introducing randomness into the demonstration selection process. It repeatedly performs the bootstrapping procedure on shuffled versions of the training set, generating multiple candidate sets of demonstrations. Each candidate is then evaluated on a validation set, and the configuration that achieves the best performance is selected. This approach improves robustness and consistency across tasks by mitigating biases from data ordering.

\end{description}

For the heuristic methods (Random, Fair, Central, and Closest), we used the validation set to tune the parameter $k$, representing the number of examples included in each prompt. Due to the stochastic nature of the random and fair sampling approaches, we executed each technique three times using different random seeds (1, 2, and 3), computing the mean accuracy and standard deviation across runs. BFS-RS was implemented using a DSPy program incorporating a Chain-of-Thought (CoT) module. The teacher and student models were configured to be identical, and execution was restricted to single-threaded mode to ensure reproducibility and eliminate potential inconsistencies due to race conditions. While this design isolates the optimizer’s effect on a single model, exploring a stronger teacher model may further improve performance and is left for future work. The optimizer requires both training and validation data internally — training data to select sets of few-shot samples, and validation data to evaluate these sets and identify the best one — so we subdivided our 20\% training set into 10\% for training and 10\% for internal validation, maintaining the original 30\% and 50\% sets for external validation and testing (final split: 10-10-30-50).

We tuned two key hyperparameters on the external validation set: $k$ (number of demonstrations per prompt) and $N$ (number of candidate sets to evaluate). 
We tested $k \in \{2, 5, 8\}$ and $N \in \{13, 30, 50, 75, 100, 150\}$. Due to single-threaded execution constraints, we limited $N$ to 150. After identifying the optimal $N$, we refined $k$ by testing values adjacent to the best initial configuration. To account for variance from training set ordering, we ran the optimizer ten times with different random shuffles (seeds 42-52), reporting mean performance and standard deviation. 

Validation experiments showed minimal deviation between validation and test performance, indicating stable generalization. The best-performing configuration for the BFS-RS optimizer corresponded to $N = 100$, with optimal $k$ values of 3 for Llama 3 and 4 for Qwen2.5. The optimal $k$ values for all methods are summarized in Table~\ref{tab:NumOfExamplesChosen}. Additional implementation details are provided in Appendix~\ref{app:bfs-rs-additional}, including the complete hyperparameter search results, conceptual information on BFS-RS, and our DSPy program, provided for reproducibility.

\begin{table}[h]
    \centering
    \caption{Optimal number of demonstrations ($k$) selected for each model (i.e., appended to the prompt) and few-shot (ICL) strategy based on validation set performance. For BFS-RS, $N$ (number of candidates explored) is also shown.}
    \label{tab:NumOfExamplesChosen}
    \begin{tabular}{llll}
        \hline
        \textbf{Model} & \textbf{ICL Strategy} & \textbf{$k$} & \textbf{Description} \\
        \hline
        Llama 3 & Random & 9 & Randomly selected \\
        Qwen2.5 & Random & 9 & Randomly selected \\
        Llama 3 & Fair & 9 & Randomly selected (stratified) \\
        Qwen2.5 & Fair & 3 & Randomly selected (stratified) \\
        Llama 3 & Central & 12 & Closest to each class centroid. \\
        Qwen2.5 & Central & 3 & Closest to each class centroid. \\
        Llama 3 & Closest & 9 & Closest to the test instance. \\
        Qwen2.5 & Closest & 9 & Closest to the test instance. \\
        Llama 3 & BFS-RS & 3 & Randomly selected, 100 candidates explored \\
        Qwen2.5 & BFS-RS & 4 & Randomly selected, 100 candidates explored \\
        
        \hline
    \end{tabular}
\end{table}

\subsection{Perturbation-based Interpretability}
To analyse how individual words or phrases influence a model's decision, we conducted experiments by systematically removing individual noun phrases from each paper’s title and abstract and observing changes in the classification output. Using spaCy \cite{Honnibal_spaCy_Industrial-strength_Natural_2020}, we extracted noun phrases as meaningful textual units and replaced each occurrence with a place holder token. The modified text was then re-classified by the model. If the prediction changed after removing a phrase, that phrase was considered important for the model’s decision. This perturbation-based approach provides an interpretable measure of which parts of the text the model relies on to determine relevance. 

\section{Results and Discussion}\label{sec:results}

Our results present a progressive investigation into whether Small Language Models (SLMs) can reliably perform nuanced classification of biomedical research papers. We begin by evaluating three-class classification performance across zero-shot and ICL strategies, then conduct class-specific analysis focusing on the RELEVANT class—our primary filtering target. Heatmap visualization reveals systematic patterns in model behavior and classification difficulty across different papers. We then examine precision–sensitivity trade-offs to assess how SLMs perform in practical deployment scenarios where literature size dictates different filtering priorities. Finally, we probe the reasoning mechanisms underlying these results, analyzing unexpected behaviors in frontier models and applying perturbation analysis to uncover the decision-making processes of SLMs.

\subsection{Three-Class Classification}\label{subsec:three-class}

Table \ref{tab:threeclass-results} presents the accuracy and other macro-averaged performance metrics for the three-class classification task (i.e., RELEVANT, SOMEWHAT RELEVANT, and IRRELEVANT). The results reveal notable performance differences across models and prompting strategies. 

\begin{table}[h]
    \centering
    \caption{Three-class classification results (macro-averaged) for zero-shot, zero-shot with rationale generation (where the model was prompted to produce an explanation alongside the predicted label); and zero-shot implementations using AdalFlow and DSPy (Predict and CoT). Results are also reported for various few-shot (in-context learning) approaches, including random few-shot (Random), fair few-shot (Fair), central examples few-shot (Central), closest examples few-shot (Closest), and bootstrap few-shot with random search (BFS-RS). The standard deviations are obtained across different random seeds (3 seeds for Random and Fair, 10 seeds for BFS-RS). The highest values in each column are shown in red, with the top performer in each category bolded. GPT-5 with “high effort” and “low effort” refers to the level of reasoning applied by the highest-tier GPT-5 reasoning model available in the API. GPT-5 Thinking Mini experiments were run through the user interface (UI), with each title-abstract uploaded in a separate temporary chat to prevent chat history from interfering with classifications. For Meerkat, predicted labels were manually extracted due to inconsistent formatting in outputs. For zero-shot (vanilla), Llama 3 and Qwen2.5 outperform GPT-5, Gemini 3 Pro Preview, and Meerkat, but clearly trail Gemini 2.5 Pro. Notably, this gap narrows with the steady improvements observed in Qwen2.5 (BFS-RS) and Llama 3 (Fair, Central).}
    \label{tab:threeclass-results}
    \resizebox{\textwidth}{!}{%
    \begin{tabular}{llccccccc}
        \toprule
        \textbf{Method} & \textbf{Model} & \textbf{Accuracy} & \textbf{Precision} & \textbf{Sensitivity} & \textbf{Specificity} & \textbf{F0.2} & \textbf{F1} & \textbf{F5} \\
        \midrule
        \multirow{4}{*}{Zero-shot (Vanilla)} & Llama 3 & 0.70 & 0.65 & 0.60 & 0.85 & 0.65 & 0.59 & 0.60 \\
        & Qwen2.5 & 0.76 & 0.56 & 0.59 & 0.87 & 0.56 & 0.56 & 0.58 \\
        & Gemini~2.5~Pro & \textbf{\textcolor{red}{0.90}} & \textbf{\textcolor{red}{0.86}} & \textbf{\textcolor{red}{0.92}} & \textbf{\textcolor{red}{0.96}} & \textbf{\textcolor{red}{0.86}} & \textbf{\textcolor{red}{0.87}} & \textbf{\textcolor{red}{0.92}} \\
        & Gemini~3~Pro~Preview & 0.72 & 0.66 & 0.62 & 0.87 & 0.66 & 0.61 & 0.62 \\
        & GPT-5 (API, high-effort) & 0.60 & 0.60 & 0.49 & 0.81 & 0.58 & 0.48 & 0.49 \\
        & GPT-5 (API, low-effort) & 0.60 & 0.63 & 0.52 & 0.82 & 0.60 & 0.50 & 0.52 \\
        & GPT-5 Thinking Mini (UI) & 0.76 & 0.79 & 0.81 & 0.91 & 0.78 & 0.72 & 0.79 \\
        & Meerkat & 0.48 & 0.55 & 0.41 & 0.75 & 0.50 & 0.37 & 0.40 \\
        \cmidrule{2-9}
        \multirow{5}{*}{Zero-shot (Rationale)} & Llama 3 & 0.68 & 0.60 & 0.56 & 0.85 & 0.59 & 0.55 & 0.55 \\
        & Qwen2.5 & 0.80 & 0.74 & 0.65 & 0.89 & 0.72 & 0.65 & 0.65 \\
        & Gemini~2.5~Pro & \textbf{\textcolor{black}{0.88}} & \textbf{\textcolor{black}{0.83}} & \textbf{\textcolor{black}{0.87}} & \textbf{\textcolor{black}{0.95}} & \textbf{\textcolor{black}{0.83}} & \textbf{\textcolor{black}{0.84}} & \textbf{\textcolor{black}{0.87}} \\
        & Gemini~3~Pro~Preview & 0.70 & 0.62 & 0.57 & 0.85 & 0.62 & 0.57 & 0.57 \\
        & GPT-5 (API, high-effort) & 0.60 & 0.63 & 0.52 & 0.82 & 0.60 & 0.50 & 0.52 \\
        & GPT-5 (API, low-effort) & 0.60 & 0.63 & 0.52 & 0.82 & 0.60 & 0.50 & 0.52 \\
        & GPT-5 Thinking Mini (UI) & 0.76 & 0.77 & 0.78 & 0.90 & 0.76 & 0.71 & 0.77 \\
        & Meerkat & 0.56 & 0.62 & 0.54 & 0.78 & 0.58 & 0.46 & 0.53 \\
        \cmidrule{2-9}
        \multirow{2}{*}{Zero-shot (AdalFlow)} & Llama 3 & \textbf{\textcolor{black}{0.74}} & \textbf{\textcolor{black}{0.70}} & \textbf{\textcolor{black}{0.67}} & \textbf{\textcolor{black}{0.87}} & \textbf{\textcolor{black}{0.69}} & \textbf{\textcolor{black}{0.65}} & \textbf{\textcolor{black}{0.66}} \\
        & Qwen2.5 & \textbf{\textcolor{black}{0.74}} & 0.64 & 0.60 & 0.86 & 0.63 & 0.60 & 0.60 \\
        \cmidrule{2-9}
        \multirow{2}{*}{Zero-shot (DSPy Predict)} & Llama 3 & 0.66 & 0.58 & 0.54 & \textbf{0.83} & 0.57 & 0.53 & 0.54 \\
        & Qwen2.5 & \textbf{0.68} & \textbf{0.61} & \textbf{0.56} & \textbf{0.83} & \textbf{0.60} & \textbf{0.55} & \textbf{0.55} \\
        \cmidrule{2-9}
        \multirow{2}{*}{Zero-shot (DSPy CoT)} & Llama 3 & 0.64 & 0.61 & 0.56 & 0.83 & 0.60 & 0.54 & 0.55 \\
        & Qwen2.5 & \textbf{0.72} & \textbf{0.69} & \textbf{0.65} & \textbf{0.86} & \textbf{0.68} & \textbf{0.63} & \textbf{0.65} \\
        \cmidrule{2-9}
        \multirow{2}{*}{Random} & Llama 3 & 0.71 $\pm$ 0.09 & \textbf{\textcolor{black}{0.68 $\pm$ 0.10}} & 0.59 $\pm$ 0.04 & 0.84 $\pm$ 0.05 & \textbf{\textcolor{black}{0.66 $\pm$ 0.08}} & \textbf{\textcolor{black}{0.58 $\pm$ 0.03}} & 0.59 $\pm$ 0.04 \\
        & Qwen2.5 & \textbf{\textcolor{black}{0.77 $\pm$ 0.01}} & 0.66 $\pm$ 0.16 & \textbf{\textcolor{black}{0.60 $\pm$ 0.02}} & \textbf{\textcolor{black}{0.86 $\pm$ 0.00}} & 0.64 $\pm$ 0.13 & \textbf{\textcolor{black}{0.58 $\pm$ 0.04}} & \textbf{\textcolor{black}{0.60 $\pm$ 0.02}} \\
        \cmidrule{2-9}
        \multirow{2}{*}{Fair} & Llama 3 & \textbf{\textcolor{black}{0.78 $\pm$ 0.06}} & 0.70 $\pm$ 0.07 & \textbf{\textcolor{black}{0.69 $\pm$ 0.08}} & \textbf{\textcolor{black}{0.90 $\pm$ 0.03}} & 0.70 $\pm$ 0.07 & \textbf{\textcolor{black}{0.69 $\pm$ 0.08}} & \textbf{\textcolor{black}{0.69 $\pm$ 0.08}} \\
        & Qwen2.5 & 0.75 $\pm$ 0.05 & \textbf{\textcolor{black}{0.74 $\pm$ 0.11}} & 0.61 $\pm$ 0.04 & 0.86 $\pm$ 0.02 & \textbf{\textcolor{black}{0.71 $\pm$ 0.09}} & 0.61 $\pm$ 0.04 & 0.61 $\pm$ 0.04 \\
        \cmidrule{2-9}
        \multirow{2}{*}{Central} & Llama 3 & \textbf{\textcolor{black}{0.82}} & \textbf{\textcolor{black}{0.72}} & \textbf{\textcolor{black}{0.67}} & \textbf{\textcolor{black}{0.90}} & \textbf{\textcolor{black}{0.71}} & \textbf{\textcolor{black}{0.66}} & \textbf{\textcolor{black}{0.67}} \\
        & Qwen2.5 & 0.74 & 0.56 & 0.57 & 0.85 & 0.55 & 0.54 & 0.57 \\
        \cmidrule{2-9}
        \multirow{2}{*}{Closest} & Llama 3 & \textbf{\textcolor{black}{0.74}} & \textbf{\textcolor{black}{0.61}} & \textbf{\textcolor{black}{0.60}} & \textbf{\textcolor{black}{0.87}} & \textbf{\textcolor{black}{0.61}} & \textbf{\textcolor{black}{0.60}} & \textbf{\textcolor{black}{0.60}} \\
        & Qwen2.5 & 0.70 & 0.54 & 0.54 & 0.83 & 0.53 & 0.51 & 0.53 \\
        \cmidrule{2-9}
        \multirow{2}{*}{BFS-RS} & Llama 3 & 0.75 $\pm$ 0.05 & 0.60 $\pm$ 0.14 & 0.58 $\pm$ 0.14 & 0.88 $\pm$ 0.02 & 0.60 $\pm$ 0.14 & 0.58 $\pm$ 0.14 & 0.58 $\pm$ 0.14 \\
        & Qwen2.5 & \textbf{\textcolor{black}{0.81 $\pm$ 0.06}} & \textbf{\textcolor{black}{0.75 $\pm$ 0.07}} & \textbf{\textcolor{black}{0.75 $\pm$ 0.09}} & \textbf{\textcolor{black}{0.91 $\pm$ 0.03}} & \textbf{\textcolor{black}{0.75 $\pm$ 0.07}} & \textbf{\textcolor{black}{0.74 $\pm$ 0.09}} & \textbf{\textcolor{black}{0.75 $\pm$ 0.09}} \\
        \bottomrule
    \end{tabular}%
    }
\end{table}

The zero-shot (vanilla) results demonstrate that small language models (SLMs) exhibit strong practical potential for this task. Notably, Llama 3 (accuracy: 0.70, $F_1$: 0.59) and Qwen2.5 (accuracy: 0.76, $F_1$: 0.56) achieved competitive performance, unexpectedly surpassing the frontier GPT-5 model in both its high- and low-effort reasoning modes (accuracy: 0.60, $F_1$: 0.48–0.50). Although Qwen2.5 attained higher accuracy than Gemini 3 Pro Preview, Llama 3 performed comparably, achieving similar accuracy and $F_1$ scores. This counterintuitive outcome—where 8B-parameter models outperform or match advanced reasoning-capable models—appears to stem from GPT-5’s strict interpretation of causal evidence and the notably poor performance of Gemini 3 Pro Preview, as further discussed in Section~\ref{subsec:llm-limitations}. Furthermore, the SLMs notably outperformed the domain-specialized Meerkat model, which underperformed despite being fine-tuned on medical reasoning datasets (accuracy: 0.48, $F_1$: 0.37), indicating limited transferability of its learned capabilities to this evaluation setting. Among all evaluated models, Gemini 2.5 Pro established the strongest baseline (accuracy: 0.90, $F_1$: 0.87), substantially outperforming every other model across all metrics. The GPT-5 Thinking Mini (UI) variant showed improved results (accuracy: 0.76, $F_1$: 0.72) compared to GPT-5’s API-based reasoning modes but still trailed behind Gemini 2.5 Pro.

The Zero-shot (Rationale) setting—where models are prompted to provide an explanatory justification alongside their predicted label—did not substantially benefit most models. Performance remained largely comparable to the vanilla zero-shot condition for the majority of models. However, two models demonstrated notable improvements: Qwen2.5 and Meerkat. Qwen2.5 exhibited the largest gains overall, with marked improvements in both accuracy and $F_1$ score. In particular, a substantial precision increase of 18\% translated into a corresponding $F_1$ gain of 9\%, highlighting enhanced decision reliability when generating rationales for its classifications . Meerkat also benefited, with accuracy rising from 0.48 to 0.56 and the $F_1$ score improving from 0.37 to 0.46, suggesting that explicit reasoning may help offset limitations in its fine-tuned domain knowledge.

The AdalFlow zero-shot implementation similarly yielded positive effects, especially for Llama 3, achieving its strongest zero-shot performance under this configuration. Accuracy and $F_1$ scores increased by approximately 5–6\%, despite only precision gains observed for Qwen2.5. In contrast, DSPy Predict and Chain-of-Thought (CoT) prompting degraded Llama 3’s performance, indicating that additional reasoning steps may introduce unnecessary noise for this model. For Qwen2.5, DSPy Predict primarily enhanced precision, while CoT generated larger overall metric gains. The fact that both rationale and CoT prompting improved Qwen2.5’s results suggests that its performance can be further strengthened when reasoning is explicitly elicited during prediction.

We next investigate whether few-shot prompting can improve performance relative to the zero-shot (vanilla) baselines. As expected, random sampling yielded only marginal gains for both models. In contrast, fair sampling provided a substantial boost for Llama 3, increasing accuracy by 8\% and $F_1$ by 10\%, while also offering a considerable precision improvement for Qwen2.5—underscoring the benefit of balancing class representation when selecting examples. Central sampling further elevated Llama 3’s performance, achieving its highest accuracy overall (0.82), compared to 0.70 in zero-shot mode, although Qwen2.5 saw minimal change under this strategy. Closest sampling offered modest increases in Llama 3’s accuracy but negatively impacted Qwen2.5’s accuracy and $F_1$, showing limited utility for either model.

The strongest improvements emerged from the Bootstrap Few-Shot with Random Search (BFS-RS) approach. Qwen2.5 showed consistent gains across all metrics, including a 19\% precision increase, a 16\% sensitivity increase, and its highest $F_1$ score (0.74) — the best among all SLM configurations. Both Llama 3 and Qwen2.5 also achieved a 5\% improvement in accuracy under BFS-RS, reflecting greater consistency in predicting all three classes. However, for Llama 3, these improvements were limited primarily to accuracy and specificity. Because BFS-RS incorporates a chain-of-thought component, its success with Qwen2.5 once again demonstrates that this model benefits from reasoning support.

Overall, these results show that few-shot effectiveness is highly dependent on both the model architecture and the sampling strategy. Qwen2.5 consistently benefits from techniques that elicit reasoning—such as Rationale prompting, DSPy CoT, and BFS-RS—often reflected in $F_1$ gains. In contrast, Llama 3 exhibits stronger improvements through simpler selection strategies like Fair and Central sampling. The following sections further unpack these observations by examining class-specific behavior and patterns.

\subsection{Class-specific Evaluation}

To gain deeper insight into our primary research question, we conducted a more focused analysis on model performance for the RELEVANT class. This class represents our main area of interest, as the goal of this study is to assess whether SLMs can serve as effective filters. In a typical use case, only papers deemed strictly relevant would progress to the next stage of screening. Therefore, we calculate additional metrics by treating RELEVANT as the positive class and combining SOMEWHAT RELEVANT and IRRELEVANT into a single negative class. It is important to note that the underlying model predictions remain three-class outputs; we simply compute evaluation metrics based on this binary grouping. These results, presented in Table~\ref{tab:binary-results}, isolate each model’s ability to identify papers that satisfy the strict relevance criteria required for downstream processing. We refer to this evaluation setup interchangeably as binary classification in the remainder of the paper

\begin{table}[h]
    \centering
    \caption{Class-specific one-vs-rest evaluation (binary classification) of the RELEVANT class. Metrics are computed by treating RELEVANT as the positive class and grouping SOMEWHAT RELEVANT + IRRELEVANT as the negative class. The results presented are for all models and all variants of zero-shot and in-context learning methods similar to Table~\ref{tab:threeclass-results}.We observe that the smaller models—Qwen2.5 (Rationale) and Llama 3 (Fair)—match the top performance of Gemini 2.5 Pro. The standard deviations are obtained across different random seeds (3 seeds for Random and Fair, 10 seeds for BFS-RS). The best score in each column is shown in red while the best performer in each category is bolded. GPT-5 with `high effort' or `low effort' refers to the level of reasoning effort applied to the most powerful GPT-5 reasoning model available through the API. GPT-5 Thinking Mini experiments were run through the user interface, with each abstract uploaded in a separate temporary chat to prevent chat history from interfering with classifications. Qwen2.5 (Rationale) and Llama 3 (Fair) match the performance of Gemini 2.5 Pro for the RELEVANT class.}
    \label{tab:binary-results}
    \resizebox{\textwidth}{!}{%
    \begin{tabular}{llccccccc}
        \hline
        \textbf{Method} & \textbf{Model} & \textbf{Accuracy} & \textbf{Precision} & \textbf{Sensitivity} & \textbf{Specificity} & \textbf{F0.2} & \textbf{F1} & \textbf{F5} \\
        \toprule        
         \multirow{4}{*}{Zero-shot (Vanilla)} & Llama 3 & 0.80 & 1.00 & 0.52 & 1.00 & 0.97 & 0.69 & 0.53 \\
        & Qwen2.5 & 0.90 & 1.00 & 0.76 & 1.00 & \textbf{\textcolor{red}{0.99}} & 0.86 & 0.77 \\
        & Gemini~2.5~Pro & \textbf{\textcolor{red}{0.92}} & \textbf{\textcolor{red}{1.00}} & {0.81} & \textbf{\textcolor{red}{1.00}} & \textbf{\textcolor{red}{0.99}} & \textbf{\textcolor{red}{0.89}} & {0.82} \\
        & Gemini~3~Pro~Preview & 0.82 & 1.00 & 0.57 & 1.00 & 0.97 & 0.73 & 0.58 \\
        & GPT-5 (API, high-effort) & 0.72 & 1.00 & 0.33 & 1.00 & 0.93 & 0.50 & 0.34 \\
        & GPT-5 (API, low-effort) & 0.70 & 1.00 & 0.29 & 1.00 & 0.91 & 0.44 & 0.29 \\
        & GPT-5 Thinking Mini (UI) & 0.78 & 1.00 & 0.48 & 1.00 & 0.96 & 0.65 & 0.49 \\
        & Meerkat & 0.68 & 0.58 & \textbf{0.90} & 0.52 & 0.58 & 0.70 & \textbf{0.89} \\
        \cmidrule{2-9}
        \multirow{5}{*}{Zero-shot (Rationale)} & Llama 3 & 0.78 & 0.92 & 0.52 & 0.97 & 0.89 & 0.67 & 0.53 \\
        & Qwen2.5 & \textbf{\textcolor{red}{0.92}} & \textbf{1.00} & 0.81 &\textbf{1.00} & \textbf{\textcolor{red}{0.99}} & \textbf{\textcolor{red}{0.89}} & 0.82 \\
        & Gemini~2.5~Pro & 0.90 & 0.94 & 0.81 & 0.97 & 0.94 & 0.87 & 0.81 \\
        & Gemini~3~Pro~Preview & 0.82 & 1.00 & 0.57 & 1.00 & 0.97 & 0.73 & 0.58 \\
        & GPT-5 (API, high-effort) & 0.70 & 1.00 & 0.29 & 1.00 & 0.91 & 0.44 & 0.29 \\
        & GPT-5 (API, low-effort) & 0.70 & 1.00 & 0.29 & 1.00 & 0.91 & 0.44 & 0.29 \\
        & GPT-5 Thinking Mini (UI) & 0.80 & 1.00 & 0.52 & 1.00 &  0.97 & 0.69 & 0.53 \\
        & Meerkat & 0.72 & 0.60 & \textbf{\textcolor{red}{1.00}} & 0.52 & 0.61 & 0.75 & \textbf{\textcolor{red}{0.98}} \\
        \cmidrule{2-9}
        \multirow{2}{*}{Zero-shot (AdalFlow)} & Llama 3 & 0.82 & \textbf{1.00} & 0.57 & \textbf{1.00} & 0.97 & 0.73 & 0.58 \\
        & Qwen2.5 & \textbf{0.86} & \textbf{1.00} & \textbf{0.67} & \textbf{1.00} & \textbf{0.98} & \textbf{0.80} & \textbf{0.68} \\
        \cmidrule{2-9}
        \multirow{2}{*}{Zero-shot (DSPy Predict)} & Llama 3 & 0.76 & 0.91 & 0.48 & 0.97 & 0.88 & 0.63 & 0.49 \\
        & Qwen2.5 & \textbf{0.80} & \textbf{1.00} & \textbf{0.52} & \textbf{1.00} & \textbf{0.97} & \textbf{0.69} & \textbf{0.53} \\
        \cmidrule{2-9}
        \multirow{2}{*}{Zero-shot (DSPy CoT)} & Llama 3 & 0.74 & 0.90 & 0.43 & 0.97 & 0.86 & 0.58 & 0.44 \\
        & Qwen2.5 & \textbf{0.80} & \textbf{1.00} & \textbf{0.52} & \textbf{1.00} & \textbf{0.97} & \textbf{0.69} & \textbf{0.53} \\
        \cmidrule{2-9}
        \multirow{2}{*}{Random} & Llama 3 & 0.77 $\pm$ 0.10 & 0.66 $\pm$ 0.09 & \textbf{\textcolor{red}{1.00 $\pm$ 0.00}} & 0.61 $\pm$ 0.16 & 0.67 $\pm$ 0.09 & 0.79 $\pm$ 0.07 & \textbf{\textcolor{red}{0.98 $\pm$ 0.01}} \\
         & Qwen2.5 & \textbf{0.89 $\pm$ 0.01} & \textbf{0.98 $\pm$ 0.03} & 0.76 $\pm$ 0.00 & \textbf{0.99 $\pm$ 0.02} & \textbf{0.97 $\pm$ 0.03} & \textbf{0.86 $\pm$ 0.01} & 0.77 $\pm$ 0.00 \\
        \cmidrule{2-9}
        \multirow{2}{*}{Fair} & Llama 3 & \textbf{0.91 $\pm$ 0.03} & \textbf{0.98 $\pm$ 0.03} & \textbf{0.81 $\pm$ 0.07} & \textbf{0.99 $\pm$ 0.02} & \textbf{0.97 $\pm$ 0.03} & \textbf{\textcolor{red}{0.89 $\pm$ 0.05}} & \textbf{0.81 $\pm$ 0.07} \\
        & Qwen2.5 & 0.86 $\pm$ 0.04 & 0.96 $\pm$ 0.03 & 0.70 $\pm$ 0.11 & 0.98 $\pm$ 0.02 & 0.94 $\pm$ 0.03 & 0.80 $\pm$ 0.07 & 0.71 $\pm$ 0.11 \\
        \cmidrule{2-9}
        \multirow{2}{*}{Central} & Llama 3 & 0.86 & 0.79 & \textbf{0.90} & 0.83 & 0.80 & \textbf{0.84} & \textbf{0.90} \\
        & Qwen2.5 & \textbf{0.88} & \textbf{1.00} & 0.71 & \textbf{1.00} & \textbf{0.98} & 0.83 & 0.72 \\
        \cmidrule{2-9}
        \multirow{2}{*}{Closest} & Llama 3 & \textbf{0.86} & 0.89 & \textbf{0.76} & 0.93 & 0.88 & \textbf{0.82} & \textbf{0.77} \\
        & Qwen2.5 & 0.84 & \textbf{1.00} & 0.62 & \textbf{1.00} & \textbf{0.98} & 0.76 & 0.63 \\
        \cmidrule{2-9}
        \multirow{2}{*}{BFS-RS} & Llama 3 & 0.83 $\pm$ 0.02 & 0.85 $\pm$ 0.06 & 0.73 $\pm$ 0.08 & 0.90 $\pm$ 0.04 & 0.84 $\pm$ 0.06 & 0.78 $\pm$ 0.04 & \textbf{0.74 $\pm$ 0.07} \\
        & Qwen2.5 & \textbf{0.88 $\pm$ 0.05} & \textbf{0.97 $\pm$ 0.05} & \textbf{0.74 $\pm$ 0.13} & \textbf{0.98 $\pm$ 0.03} & \textbf{0.95 $\pm$ 0.04} & \textbf{0.83 $\pm$ 0.08} & \textbf{0.74 $\pm$ 0.13} \\
        \bottomrule
    \end{tabular}%
    }
\end{table}

When isolating performance on the RELEVANT class, all models exhibit strong zero-shot (vanilla) baseline results. Among the SLMs, Qwen2.5 stands out, achieving an accuracy of 0.90 and an $F_1$ score of 0.86. Its performance is second only to the state-of-the-art Gemini 2.5 Pro (accuracy: 0.92, $F_1$: 0.89), demonstrating that an 8B-parameter model can nearly match the best frontier model on the task of filtering strictly relevant papers. Llama 3 performs comparably to Gemini 3 Pro Preview, with both models achieving perfect precision but reduced sensitivity—indicating highly cautious inclusion criteria. A similar pattern is observed in the GPT-5 variants, which also achieve perfect precision yet struggle considerably with recall (sensitivity).

A particularly noteworthy finding is the performance shift of GPT-5 Thinking Mini: although it ranked as one of the strongest models under the multiclass evaluation in Table~\ref{tab:threeclass-results}, it appears to struggle in accurately distinguishing the RELEVANT class, suggesting that its earlier performance benefits may have come from strong predictions on the non-target classes. More broadly, nearly all models (except Meerkat) exhibit perfect precision and specificity, reflecting extremely low false-positive rates—i.e., when they label a paper as relevant, they are almost always correct. However, this conservative behavior comes with the expected trade-off: diminished sensitivity due to under-prediction of truly relevant papers. In contrast, the domain-tuned Meerkat model presents the opposite precision–recall balance. While its precision is relatively low (0.58), its sensitivity reaches 0.90—the highest among all models—indicating a highly liberal classification strategy that prioritizes capturing as many relevant papers as possible, even at the cost of increased false positives.

For the RELEVANT class, the Zero-shot (Rationale) setting benefited only Qwen2.5 and Meerkat—mirroring trends from the three-class evaluation. Under this configuration, Meerkat achieved perfect sensitivity, while Qwen2.5 matched the performance of the top-ranked Gemini 2.5 Pro. This is a striking result: with the addition of explicit reasoning, Qwen2.5 becomes equally effective as the state-of-the-art frontier model at identifying strictly relevant papers. However, this also suggests that Qwen2.5’s weaker performance in the multiclass results is likely due to misclassifications in the non-target classes. In contrast, the AdalFlow and DSPy implementations did not yield improvements for either Llama 3 or Qwen2.5—particularly surprising for Qwen2.5, given that CoT had previously been beneficial.

Few-shot prompting revealed performance patterns that closely align with the three-class results. With random sampling, Llama 3 traded its previously perfect precision for perfect sensitivity, while Qwen2.5 remained largely unchanged. This behavior in Llama 3 was unexpected: it achieved perfect sensitivity for the RELEVANT class, meaning it readily classifies papers as RELEVANT but at the cost of misclassifications in other classes. Fair sampling led to substantial gains for Llama 3, achieving 0.91 accuracy and an $F_1$ score of 0.89—matching the highest performance observed for both Gemini 2.5 Pro and Qwen2.5 under Zero-shot (Rationale). Central, Closest, and BFS-RS sampling strategies also improved Llama 3’s RELEVANT class performance relative to vanilla zero-shot, whereas Qwen2.5 showed minimal change or even slight declines across these methods, including BFS-RS.

Together, these findings reveal a clear divide in how the two SLMs leverage contextual information. Llama 3 benefits substantially from in-context learning strategies when identifying strictly relevant papers, while Qwen2.5 reaches near-optimal performance with its zero-shot configuration alone. Importantly, both models are capable of achieving performance on par with Gemini 2.5 Pro for the RELEVANT class—demonstrating that small models can serve as highly effective filters for critical literature screening tasks.

\subsection{Paper-Specific Classification Analysis}

\begin{figure}[H]
    \centering
    \includegraphics[width=1.0\linewidth]{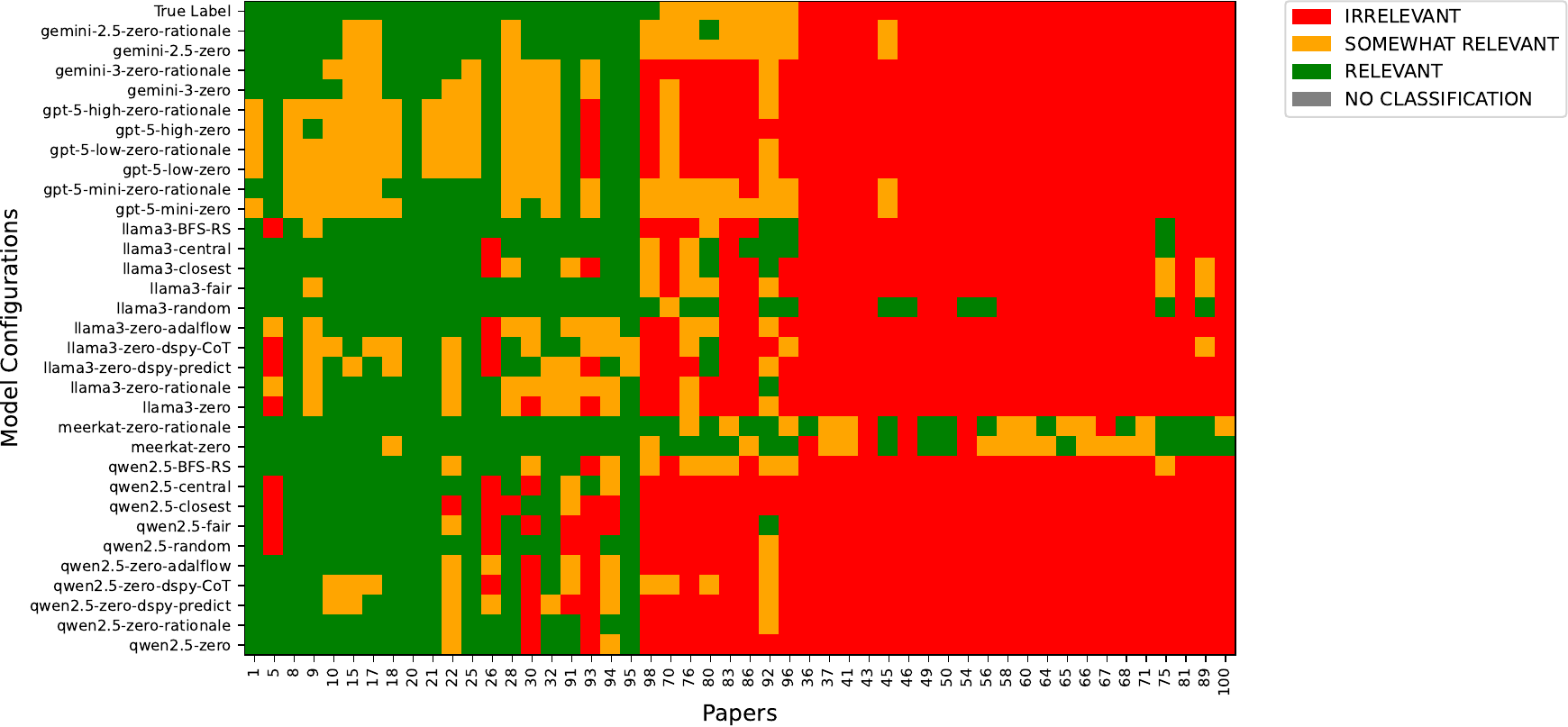}
    \caption{Heatmap performance comparison of zero-shot, zero-shot variants and ICL (few-shot) techniques over the test set examples for 32 model-technique configurations. The top row gives the ground truth label with relevance classifications color-coded as: RELEVANT (green), SOMEWHAT RELEVANT (yellow), IRRELEVANT (red), and NO CLASSIFICATION (gray). The classification labels were aggregated for techniques with multiple runs (Random, Fair, BFS-RS). We see that most models struggled to correctly predict the SOMEWHAT RELEVANT label accurately while retaining good performance on the RELEVANT class. GPT-5 was uniformly conservative, mostly classifying RELEVANT papers as SOMEWHAT RELEVANT. In contrast GPT-5-mini performed better. Note that all models successfully predicted a label — none produced a No Classification result. Models with the ‘-zero’ suffix denote the vanilla zero-shot setting.}
    \Description[Heatmap of model predictions versus true labels]{The figure shows a heatmap comparing predicted relevance classifications for each model configuration against the true labels across all test papers. The top row displays the ground truth sequence of labels, while each subsequent row corresponds to a different model and prompting configuration. Each column represents one paper, aligned across all models. Green indicates relevant, yellow indicates somewhat relevant, red indicates irrelevant, and gray indicates no classification. The heatmap reveals that errors are concentrated around the somewhat relevant class, where many models fluctuate between relevant and irrelevant. Several models show systematic over- or under-prediction trends, visible as large contiguous color blocks. The irrelevant class appears consistently stable across models, while individual columns highlight particularly difficult papers that multiple models misclassify in similar ways.}

    \label{fig:heatmap}
\end{figure}

Figure~\ref{fig:heatmap} visualizes the classifications of each model configuration across all test-set papers, allowing for a transparent comparison of how different prompting strategies influence model predictions relative to the true labels for the 50 papers. Notably, Gemini 2.5 Pro demonstrates outstanding performance—its heatmap is nearly indistinguishable from the true labels, indicating highly accurate predictions across almost all papers. In addition to this strong overall performance, the heatmap reveals several other key observations, outlined below.

\subsubsection{The SOMEWHAT RELEVANT Class is the Most Challenging}
The heatmap highlights that the SOMEWHAT RELEVANT class is the primary source of model confusion. When the true label is SOMEWHAT RELEVANT (yellow), many models alternate between predicting RELEVANT (green) and IRRELEVANT (red), illustrating difficulty in distinguishing borderline cases. Only Gemini 2.5 Pro and GPT-5 Mini show consistently strong performance on this class, while Gemini 3 Pro Preview, Meerkat, and the vanilla SLMs struggle considerably. However, we observe meaningful gains from applying in-context learning techniques to SLMs. In particular, Qwen2.5 with BFS-RS exhibits the strongest improvements, followed by Llama 3 with Fair sampling and Qwen2.5 with DSPy CoT. These results suggest that Chain of Thought prompting helps Qwen2.5 better capture the nuanced distinctions that define SOMEWHAT RELEVANT papers. Among the SLMs, Qwen2.5 ultimately achieves the highest performance on this challenging class. Collectively, these findings highlight the inherent ambiguity of the SOMEWHAT RELEVANT class and reinforce the nuanced nature of the classification task itself.

\subsubsection{Frontier LLMs have strict requirements for causality}

GPT-5, and to a certain extent Gemini 3 Pro Preview, exhibit very strict interpretation of causal relevance criteria, as is clearly visible in the heatmap with the large yellow blocks in the RELEVANT, green region — indicating systematic underclassification of relevant papers. All GPT-5 variants we evaluated, including the Mini version, exhibit this behavior to some extent. This pattern helps explain the earlier discrepancy where GPT-5 Mini performed strongly in the multiclass evaluation, yet struggled specifically on the RELEVANT class. This conservatism cascades into misclassification of the SOMEWHAT RELEVANT class by some models as well: when papers that should be RELEVANT are downgraded to SOMEWHAT RELEVANT, those that are SOMEWHAT RELEVANT are often incorrectly classified as IRRELEVANT.  We also observe that Gemini 3 Pro Preview introduces more SOMEWHAT RELEVANT predictions for truly RELEVANT papers than Gemini 2.5 Pro, suggesting a similar over-cautious, overly fussy bias. The potential causes and implications of these behaviors are explored further in Section~\ref{subsec:llm-limitations}.

\subsubsection{Over-Prediction of the RELEVANT Class and the Source of High Sensitivity}
The heatmap clarifies why Meerkat and Llama 3 with Random ICL exhibit high sensitivity for the RELEVANT class: both models tend to over-predict RELEVANT labels. This behavior increases sensitivity by capturing more truly relevant papers, but comes at the cost of misclassifying many SOMEWHAT RELEVANT and IRRELEVANT papers as RELEVANT (i.e., reduced precision). In other words, these models prioritize minimizing false negatives—even if that leads to a higher number of false positives.

\subsubsection{Model-specific patterns}

We observe several important error patterns across both SLMs and the larger models. First, certain papers appear inherently difficult for the SLMs to classify correctly, leading to precision–sensitivity trade-offs when attempting to improve performance. For instance, Llama 3 gains substantial improvements through ICL techniques; however, this comes at the cost of increased false positives. While zero-shot variants of Llama 3 makes almost no errors on the IRRELEVANT class, post-ICL it begins to misclassify papers such as 75 and 89 as RELEVANT or SOMEWHAT RELEVANT, resulting in a loss of its earlier perfect precision. Additionally, Llama 3 is noticeably more prone than Qwen2.5 to upgrading SOMEWHAT RELEVANT papers into the RELEVANT class, reflecting a more flexible classification boundary after ICL is applied.

Qwen 2.5, in contrast, is remarkably reliable in identifying IRRELEVANT papers across all configurations. However, its behavior in the RELEVANT region of the heatmap reveals why its performance sometimes declines with ICL. Under zero-shot prompting, Qwen2.5 consistently mislabels a small subset of RELEVANT papers—such as papers 30 and 93—as IRRELEVANT. When ICL techniques are introduced (except BFS-RS), it begins misclassifying additional RELEVANT papers (e.g., paper 5) into the negative class, which explains the reduced sensitivity and $F_1$ score observed under those setups. Finally, we note that the IRRELEVANT class is consistently the easiest for all models to classify, which aligns with expectations given that several papers in this category are entirely unrelated to MMTV-like virus and breast cancer, providing clearer evidence for exclusion.

\subsubsection{Difficult papers for classification}
Some papers present consistent challenges for SLMs in the RELEVANT class, with models frequently predicting SOMEWHAT RELEVANT or IRRELEVANT instead. Papers 5, 26, 30, 93, and 98 are particularly challenging, with several models consistently misclassifying them despite their true label being RELEVANT. ICL strategies help some models recover, but improvements remain inconsistent across papers. Notably, Paper 98 proved exceptionally difficult even for frontier models, with only Meerkat and Llama 3 (Random ICL) classifying it correctly. We suspect these correct predictions reflect a bias toward the RELEVANT class rather than genuine understanding, motivating a closer inspection.

Paper 98 investigates mammary cancer in mice using an MMTV-neu transgene, which combines the MMTV promoter with the HER2/neu oncogene, and compares tumor histology with human breast cancers exhibiting similar protein levels. This design meets the criteria for a relevant classification, as it illustrates how an MMTV-like sequence upstream of a proto-oncogene could plausibly contribute to cancer. Yet most models misclassified the paper: Gemini 3 Pro Preview, GPT-5 (high and low reasoning), Llama 3, and Qwen2.5 labeled it irrelevant, while Gemini 2.5 Pro rated it somewhat relevant. Some argued the MMTV promoter was “merely a tool” for inducing HER2/neu expression and did not address viral causality, overlooking that the study explicitly involves MMTV and its oncogenic context. Only Meerkat and Random Llama 3 classified it correctly, though their reasoning was vague and likely biased toward relevance. Overall, this case highlights a rare borderline scenario where even experts—and most models—might reasonably disagree due to the nuanced biological implications of the MMTV-neu construct.

\subsection{Precision-Sensitivity Analysis and Practical Implications}\label{subsection:prec-sens}

For the AI co-scientist pipeline, our primary objective is to ensure that RELEVANT papers are not filtered out prematurely. However, this depends on the size of the literature for a given microbe–cancer pair:
\begin{itemize}
    \item \textbf{Case 1 — Large Search Space:} When the literature search returns a large number of papers, misclassifying IRRELEVANT papers as RELEVANT becomes costly, as it increases both downstream processing burden and financial requirements. Missing a small number of RELEVANT papers is less critical in this context because the remaining literature is likely to capture similar knowledge. Therefore, only papers predicted as RELEVANT are passed forward. In this scenario, we prioritize \textbf{perfect precision} while maintaining \textbf{high sensitivity}.
    \item \textbf{Case 2 — Small Search Space:} When few papers exist for a particular microbe–cancer pair, missing even a single RELEVANT paper poses a significant risk to knowledge discovery. Passing a handful of IRRELEVANT papers forward has minimal cost implications, as the downstream LLM filtering step can easily remove them. Accordingly, both RELEVANT and SOMEWHAT RELEVANT papers are passed forward. Here, we prioritize \textbf{high precision} while ensuring \textbf{perfect sensitivity}.
\end{itemize}

\begin{figure}[H]
    \centering
    \includegraphics[width=0.95\linewidth]{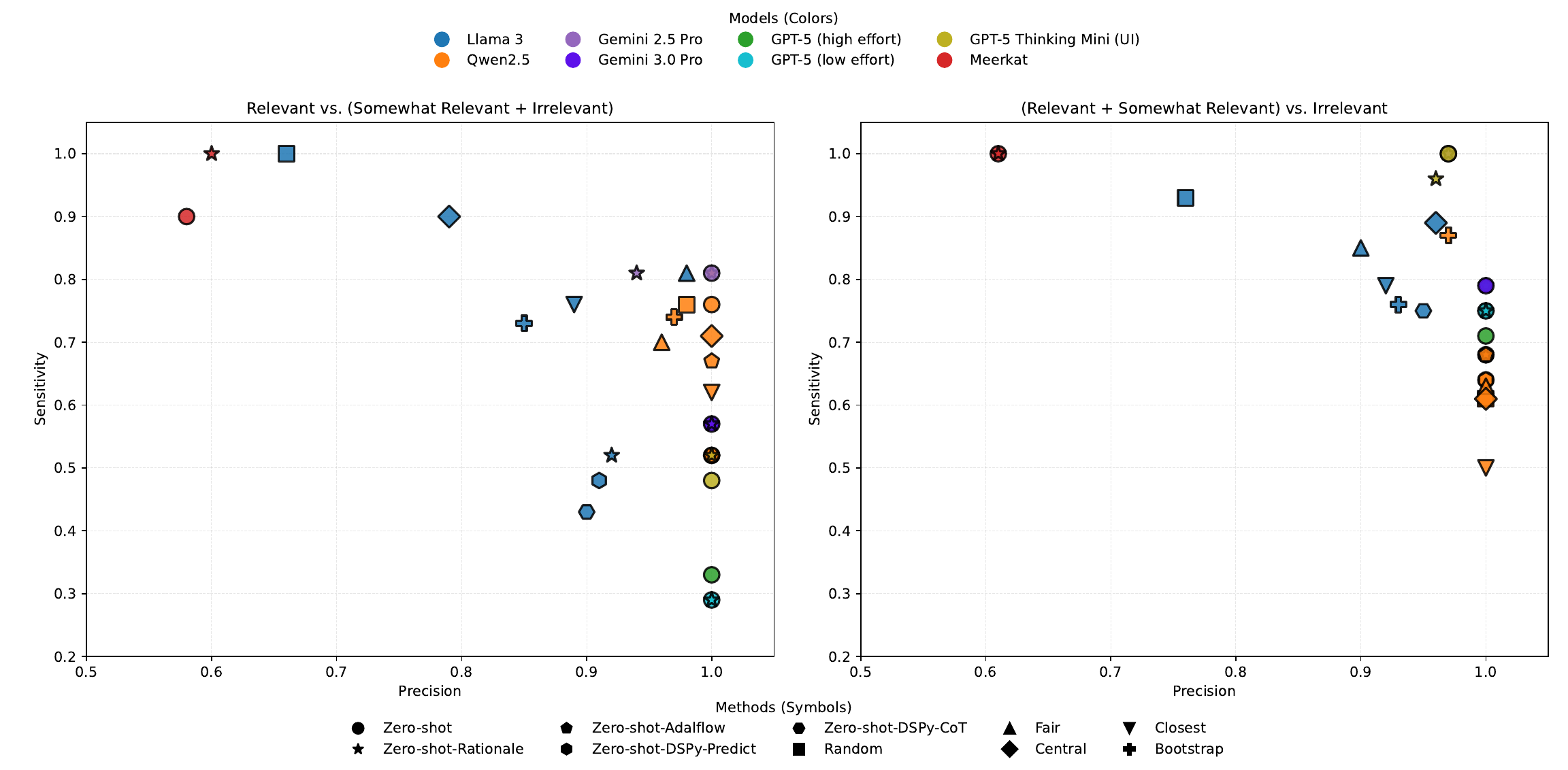}
    \caption{Comparison of sensitivity (recall) vs. precision across all models (shown by symbol colour) and in-context learning modes (symbol shape). {\bf Left:} Binary classification performance with RELEVANT as the positive class and SOMEWHAT RELEVANT and IRRELEVANT grouped as the negative class. {\bf Right:} Binary classification performance with RELEVANT and SOMEWHAT RELEVANT grouped as the positive class and IRRELEVANT as the negative class. These settings reflect different priorities that may emerge depending on the number of retrieved papers for a Microbe–Cancer Pair (MCP). A small number of returned papers suggests prioritizing sensitivity and combining RELEVANT and SOMEWHAT RELEVANT papers ({\bf case 2}) to pass forward (right panel), while a large number of returned papers may prioritise precision and consider only RELEVANT papers ({\bf case 1}) to pass forward (left panel). We observe a clear precision-sensitivity trade-off for LLama 3 (Left panel), with multiple models achieving near-perfect precision. Qwen2.5 Rationale (orange star, under light purple circle) and Llama 3 Fair (blue triangle) closely match Gemini 2.5 Pro in this case. On the other hand (right panel), BFS-RS for Qwen2.5 (orange cross) and Llama 3 Central (blue diamond) perform exceptionally well as they approach the performance of the best commercial frontier models (Gemini 2.5 Pro and GPT-5 Thinking Mini).}
    \Description[Precision–sensitivity trade-offs across models and filtering strategies]{The figure contains two scatter plots showing the relationship between precision and sensitivity for multiple model and in-context learning configurations. Each point represents one configuration, with color indicating the model family and symbol indicating the prompting or sampling method. The left panel treats only relevant papers as the positive class, while the right panel groups relevant and somewhat relevant papers as positive. Across both panels, different model families occupy distinct regions of the plots, illustrating systematic trade-offs between precision and sensitivity. In the left panel, several configurations cluster near perfect precision with varying sensitivity, while others achieve perfect sensitivity at reduced precision. In the right panel, more configurations shift toward the upper-right region, reflecting higher sensitivity when the positive class is expanded. The distribution of points highlights how different modeling and prompting choices lead to distinct operating regimes depending on whether precision or sensitivity is prioritized.}
    \label{fig:precisionSensTradeOff}
    \end{figure}

In summary, the prioritization strategy must adapt to the scale of the literature retrieved. When a large number of papers are returned, precision becomes the primary objective to avoid overwhelming the downstream system with irrelevant studies. Conversely, when only a small number of papers are available, sensitivity is paramount to ensure that no potentially valuable evidence is excluded from further analysis. Figure \ref{fig:precisionSensTradeOff} illustrates the precision–sensitivity dynamics across two evaluation settings: Case 1 where only RELEVANT class is treated as positive (left), and Case 2 where RELEVANT + SOMEWHAT RELEVANT classes are grouped as positive (right).

For Case 1 (left panel), both Llama 3 and Qwen2.5 exhibit an inverse relationship between precision and sensitivity: methods with higher precision generally show lower sensitivity, and vice versa. This trade-off is particularly pronounced for Llama 3, which spans nearly the full precision–sensitivity spectrum—from perfect precision with moderate sensitivity in vanilla zero-shot mode (1.0 precision, 0.52 sensitivity) to perfect sensitivity with reduced precision under random sampling (1.0 sensitivity, 0.66 precision). In contrast, Qwen2.5 remains tightly clustered near the perfect-precision region, showing limited variability across ICL methods. While Llama 3 drifts away from this boundary as more ICL strategies are applied, most model configurations cluster near perfect precision, reflecting the previously noted conservative behavior that minimizes false positives. Under this scenario, the optimal operating point prioritizes perfect precision with high sensitivity. Llama 3 (Central) provides a good balance between precision and sensitivity, while most configurations of Qwen2.5, including zero-shot, reside in the perfect-precision–decent-sensitivity region, highlighting its strength on the RELEVANT class for this task. Gemini 2.5 Pro delivers the strongest overall performance, achieving perfect precision with 0.81 sensitivity. However, both Qwen2.5 (Rationale) and Llama 3 (Fair sampling) achieve comparable results, demonstrating that SLMs can serve as effective filters in Case 1.

For Case 2, we observe a similar precision–sensitivity trade-off, but with systematically higher sensitivities across many configurations. Llama 3 again spans a broad operational range under different ICL strategies, while Qwen2.5 remains unchanged except under BFS-RS. In this setting, the optimal configuration reverses the priority: high precision with perfect sensitivity. As expected, more configurations occupy the upper right region of the plot compared to Case 1, since collapsing SOMEWHAT RELEVANT and RELEVANT into a single pass-forward class increases the likelihood of capturing all relevant papers while reducing misclassification of IRRELEVANT papers, which are easier to distinguish, as noted previously. Llama 3 configurations generally reside in this upper region, whereas Qwen2.5 configurations exhibit lower performance due to a higher tendency to misclassify IRRELEVANT papers. Gemini 2.5 Pro and GPT-5 Mini both achieve perfect sensitivity with very high precision. Meerkat also attains perfect sensitivity, though at a lower precision (0.60). Importantly, Qwen2.5 BFS-RS and Llama 3 with ICL strategies perform strongly, approaching frontier-model precision while maintaining sensitivities near 0.90 — confirming that SLMs can also serve as effective filters in Case 2. Notably, although Meerkat trails the strongest models in precision, its perfect sensitivity is impressive given its smaller 7B parameter scale — suggesting that targeted ICL improvements may yield substantial gains. We leave this direction to future work.

Overall, this analysis demonstrates that different model configurations exhibit distinct precision–sensitivity trade-offs, allowing strategic selection based on the literature characteristics for each microbe–cancer pair. These patterns have important implications for deployment within the AI co-scientist filtering pipeline: high-precision configurations reduce downstream review workload, while high-sensitivity setups maximize discovery potential by minimizing the risk of discarding relevant evidence. The strong performance of SLMs enhanced with ICL makes them particularly well-suited for real-world use, balancing coverage and reliability. In the following subsections, we examine the limitations of frontier LLMs and use perturbation-based analysis to further illuminate SLM decision-making behavior.

\subsection{Performance Limitations and Insights from Frontier LLMs}\label{subsec:llm-limitations}

The unexpectedly weak performance of frontier LLMs, including Gemini 3 Pro Preview and GPT-5, motivated a deeper examination of the frontier models’ underlying reasoning processes. Their inconsistent classifications raised concerns about the reliability of the expert-labeled dataset and prompted further analysis. To address these issues, we closely investigate the reasoning produced by these frontier models to better understand the basis of their predictions.

\subsubsection{Gemini Zero-Shot Performance}

Despite achieving the highest overall metrics in the three-class classification task, Gemini 2.5 Pro, being a state-of-the-art model, did not achieve near-perfect performance across some measures, revealing the difficulty of the task itself. Inspection of Gemini 2.5 Pro’s reasoning indicated that most errors originated from the model rather than annotation noise. While Gemini 2.5 Pro occasionally displayed nuanced domain reasoning—prompting re-evaluation of two papers by experts—it frequently produced inconsistent or superficial judgments. For example, the model correctly classified several correlational studies on MMTV and breast cancer yet misclassified others with nearly identical designs, focusing on authorial framing (“causal argument”) rather than the actual study design. Similar inconsistencies appeared when abstracts mentioned MMTV without breast-cancer context or when it over-generalized findings from unrelated cancers. 

Gemini 3 Pro Preview exhibited similar shortcomings to Gemini 2.5 Pro but performed more poorly overall, particularly for relevant and somewhat relevant papers. Its outputs were marked by inconsistency: correlational studies were alternately deemed sufficient evidence for causality or dismissed as merely associational, and papers mentioning both MMTV and breast cancer but not investigating viral etiology were often misclassified as irrelevant. For these, Gemini 3 Pro Preview reasoned that the MMTV promoter was used only as a “genetic tool” for investigating other relationships, overlooking that such studies meet the somewhat relevant criterion of mentioning both entities without exploring causality. It also relied heavily on authorial claims for relevant classifications, citing these as “causal arguments” rather than relying on study design or results. Unlike Gemini 2.5 Pro, Gemini 3 Pro Preview misinterpreted the abstract of “Multihormone Regulation of MMTV-LTR in Transfected T-47-D Human Breast Cancer Cells” \cite{Glover1989}, concluding that the paper examined the effects of MMTV’s LTR region on hormone regulation when in reality the paper investigated the inverse (i.e., the effects of steroid hormones on MMTV expression), leading to a misclassification of somewhat relevant instead of relevant. While Gemini 3 Pro Preview avoided the plausibility over-generalization errors seen in Gemini 2.5 Pro, it failed to generalize appropriately when findings from other viruses in breast cancer could have been applied, underscoring persistent weaknesses in inferential reasoning.

\subsubsection{GPT-5 Zero-Shot Performance}

Both GPT-5 reasoning variants performed below expectations in the three-class task, with systematic under-classification of relevant and somewhat relevant work. In contrast to Gemini 2.5 Pro, both variants generally applied the classification criteria strictly, favoring explicit causal statements and overlooking correlational evidence, studies demonstrating viral oncogenicity in animal models, and those providing mechanistic insights that lacked overt causal language, with the exception of correlational studies with negative findings, which both variants correctly classified as relevant, indicating an overemphasis on results rather than study design. While it could be argued, as GPT-5 did, that correlational studies do not produce true causal evidence, some of these studies would still fulfill the other relevant criterion which calls for a theoretical model for how HMTV/MMTV-like virus might cause breast cancer, irrespective of whether there is experimental evidence for it. Yet, GPT-5 stated that these papers did not provide theoretical modeling ("no explicit model") of this relationship, implying that it interpreted this criterion as requiring literal modeling of the relationship, while the criterion itself very clearly does not call for this. 

These findings illustrate that even frontier reasoning models remain brittle when confronted with nuanced, domain-specific criteria.

\subsection{Understanding SLMs' Decision-Making Through Perturbation Analysis}

To investigate which textual features influence SLM classification decisions, we performed a perturbation analysis on Qwen2.5 and Llama 3’s zero-shot (vanilla) predictions by systematically removing words or phrases and observing the resulting classification changes across all papers in the dataset. This approach reveals the model’s sensitivity to specific terms within titles and abstracts, illustrating both the strengths and brittleness of current SLMs in complex classification tasks.

\subsubsection{Models Rely on Scientifically Appropriate Cues}

\begin{figure}[ht]
    \centering
    \includegraphics[width=1\linewidth]{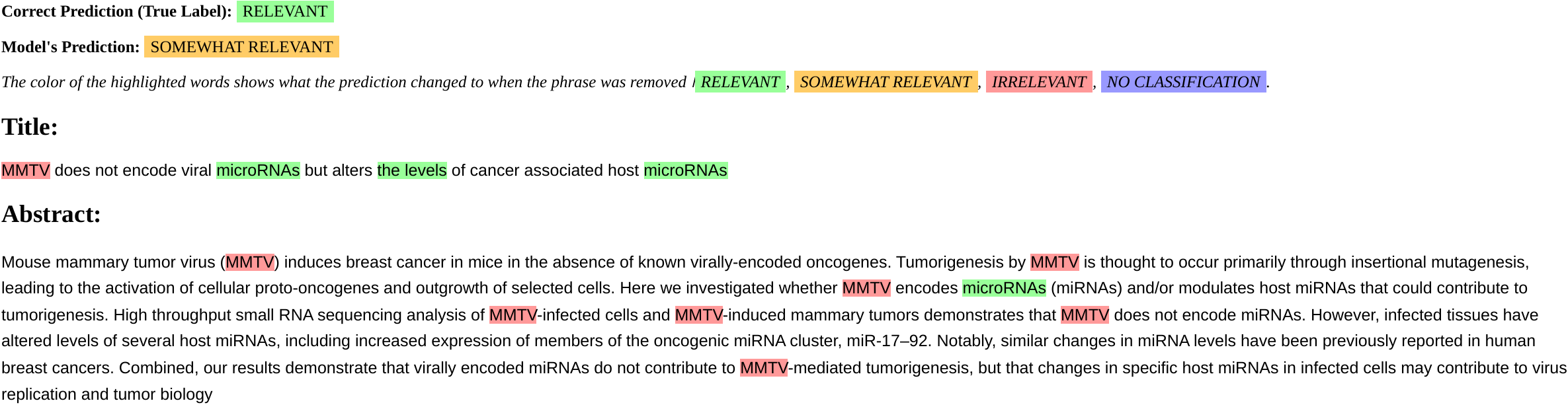}
    \caption{Word removal analysis for Paper 91 with Qwen2.5. The highlighted words represent those removed during analysis, with each color denoting the new class assigned after removal (green for \textsc{RELEVANT}, orange for \textsc{SOMEWHAT RELEVANT}, red for \textsc{IRRELEVANT}, and purple for \textsc{NO CLASSIFICATION}). As expected, removing the key term “MMTV” decreased relevance scores, but removing other relevant terms (e.g.“microRNAs”) instead unexpectedly increased relevance}
    \label{fig:P91_qwen}
    \Description[Word removal analysis showing model sensitivity to biomedical terms in Paper 91]{Word removal analysis for Paper 91 with Qwen2.5 showing a title and abstract with color-coded highlights. Each highlighted word or phrase indicates the classification change when removed: green for relevant, orange for somewhat relevant, red for irrelevant, and purple for no classification. Removing the key virus term “MMTV” decreased relevance scores as expected, but removing other scientifically meaningful terms like “microRNAs” paradoxically increased relevance scores, revealing inconsistent model reasoning despite correct keyword identification.}
\end{figure}

Analysis of correctly classified papers revealed that Qwen2.5 often relies on scientifically meaningful cues. For example, in Paper 91 (Figure~\ref{fig:P91_qwen}), removing the key term "MMTV"—the primary virus of interest—caused the model to decrease its relevance classification, demonstrating accurate recognition of a central biomedical concept. This behavior indicates that the model can correctly identify domain-specific terminology central to the classification task. Similar behaviours were observed across other papers as well.

\subsubsection{Models Are Also Susceptible to Spurious Features}
\begin{figure}[ht]
    \centering
    \includegraphics[width=1\linewidth]{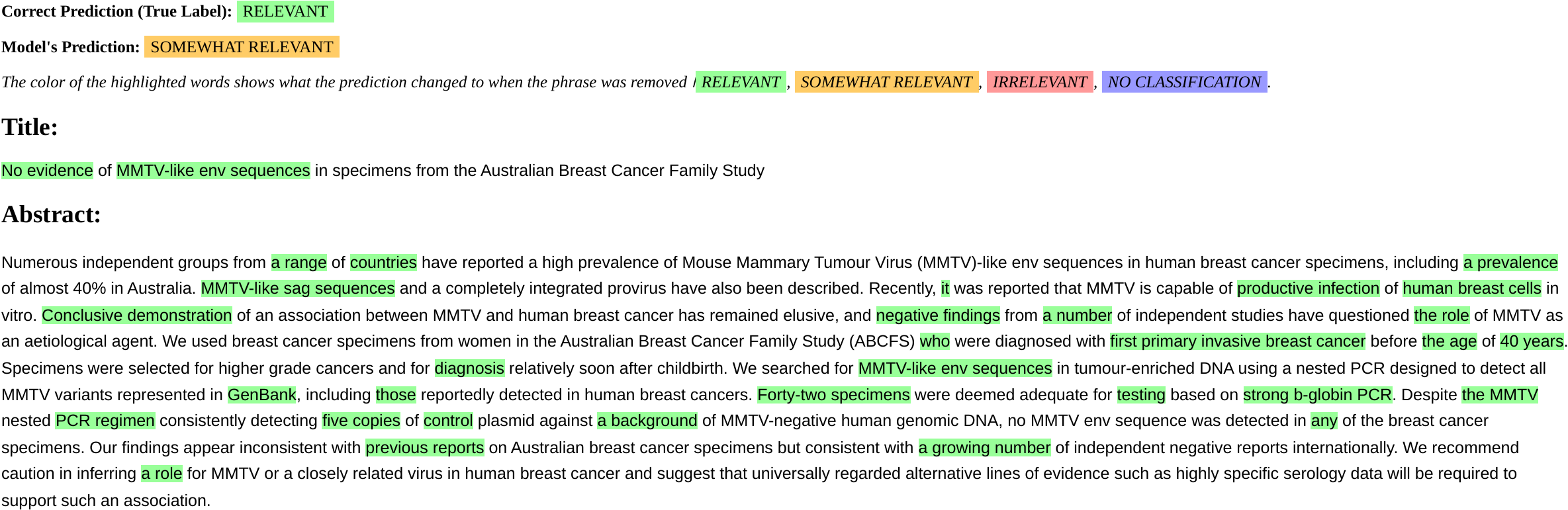}
    \caption{Word removal analysis for Paper 26 with Qwen2.5. he highlighted words represent those removed during analysis, with each color denoting the new class assigned after removal (green for \textsc{RELEVANT}, orange for \textsc{SOMEWHAT RELEVANT}, red for \textsc{IRRELEVANT}, and purple for \textsc{NO CLASSIFICATION}). Removing key terms (e.g., “MMTV-like env sequences”) unexpectedly increased relevance scores, and removing irrelevant words (e.g., “countries”) produced the same effect. This highlights the model’s inconsistent and counterintuitive sensitivity to input terms.}
    \label{fig:P26_qwen}
    \Description[Word removal analysis demonstrating spurious feature sensitivity in Paper 26]{Word removal analysis for Paper 26 with Qwen2.5 showing a title and abstract with color-coded highlights. Each highlighted word or phrase indicates the classification change when removed: green for relevant, orange for somewhat relevant, red for irrelevant, and purple for no classification. Removing critical phrases like “MMTV-like env sequences” unexpectedly increased relevance scores instead of decreasing them. Similarly, removing clearly irrelevant terms such as “countries”, “it” and “40 years” also increased relevance, demonstrating the model's reliance on spurious textual correlations rather than semantic understanding.}
\end{figure}

Despite these strengths, the analysis also revealed notable brittleness in the model’s reasoning. Paper 26 (Figure \ref{fig:P26_qwen}) exemplifies this issue: removing the critical phrase “MMTV-like env sequences”—directly relevant to the research question—caused the model to increase its relevance rating, contrary to expectations. The same pattern was observed when removing other key terms such as “MMTV-like sag sequences,” “human breast cells,” and “first primary invasive breast cancer.” Intuitively, eliminating key topic-related phrases should make the paper appear less relevant, not more. Even more strikingly, removing clearly irrelevant terms (e.g., “countries,” “it,” “40 years,” “testing,” and “control”) also increased the model’s relevance rating. This pattern suggests that the model’s predictions are driven in part by shallow textual correlations rather than genuine semantic understanding.

Paper 91 (Figure~\ref{fig:P91_qwen}) shows a similar inconsistency. While removing "MMTV" correctly decreased relevance as noted above, removing other relevant terms like "the levels" or "microRNAs"—clearly meaningful biomedical concepts central to the investigation—paradoxically  \emph{increased} the relevance score. This inconsistency reveals that the model does not reliably distinguish between scientifically meaningful terms and contextual noise. Together, these examples highlight a systematic issue: although the model's aggregate performance may appear strong, the underlying reasoning is not consistently grounded in nuanced scientific understanding. 

Finally, analysis of Llama 3 revealed similar patterns to those observed in Qwen2.5, suggesting that these behaviors might reflect general properties of current 8B-parameter models rather than model-specific limitations. Both systems combine reasonable keyword identification with vulnerability to spurious textual cues, indicating a broader tendency among small language models in complex domain-specific tasks. Analysis examples for Llama 3 are provided in Appendix~\ref{app:perturbation}.

\subsubsection{Implications for SLM Deployment in Scientific Workflows}

This dual character—correct identification of domain-relevant keywords alongside sensitivity to spurious features—has important implications for deploying SLMs in scientific screening pipelines. The perturbation analysis helps explain an apparent paradox in our results: despite achieving competitive performance, the models' decision-making process reveals gaps in reasoning capability. This outcome is not unexpected given the nuanced nature of the task and its difficulty even for state-of-the-art frontier models, as discussed in Section~\ref{subsec:llm-limitations}.

These observations support our proposed role for SLMs within AI co-scientist pipelines: they function effectively as first-pass filters that narrow large literature corpora to manageable subsets, but they should not serve as autonomous decision-makers for subsequent stages of the co-scientist pipeline. The quantitative results highlight their practical utility in the filtering stage, while the qualitative analysis explains why frontier reasoning models—or human oversight—remain essential for downstream stages and final assessment. Notably, frontier LLMs do not exhibit substantial advantages over SLMs for this filtering task. They show their own limitations, and SLMs can match their performance while being far more cost-efficient. This makes SLMs a natural and practical choice for large-scale literature screening, especially when computational efficiency and scalability are critical.

\section{Limitations}\label{sec:limitations}

A key limitation of this study is the relatively small dataset of 100 papers. Constructing it required extensive manual review and annotation by domain experts to ensure accurate relevance labeling against our nuanced criteria, resulting in a high-quality collection despite its modest size. Nevertheless, the dataset's scale limits our ability to test generalization of our findings. For complex optimizers such as BFS-RS, a larger dataset would enable more effective sampling of few-shot examples. Despite this, our models successfully identified strong examples from the training set, tuned hyperparameters for generalization on the validation set, and achieved robust results on the test set which would be good to confirm with larger datasets.  

Our study was designed to use titles and abstracts of papers, rather than full-text articles. As a result, all analyses and model performance evaluations are constrained to the information available in these sections. The performance of language models—particularly in identifying causal evidence for microbe–cancer pairs (MCPs)—may differ when full papers are considered, where additional context, methods, and results could provide richer evidence. Consequently, the reported performance of both small and frontier LLMs is specific to this dataset and may underestimate their true capabilities. Future work should investigate model performance on full-text articles to provide a more comprehensive assessment.

The specific small language models (SLMs) we selected—Qwen2.5 and Llama 3—are not necessarily representative of all current-generation SLMs. We chose these models primarily for their favorable timing benchmarks, faster execution, and strong reasoning performance. We acknowledge that other models of similar scale (around 8B parameters) may behave differently and could provide additional insights. Our goal was not to claim that nuanced reasoning is generalizable across all SLMs, but rather to demonstrate that the selected models (SLMs) can perform this task effectively and achieve performance comparable to larger, frontier models. This naturally motivates future research to investigate whether tasks like these are generalizable across different SLMs.

While this work is motivated by the broader vision of an AI co-scientist capable of discovering, filtering, and deeply understanding scientific literature, our implementation and evaluation focus exclusively on the filtering stage of this pipeline. Specifically, we investigate whether small language models (SLMs) can serve as effective filtering agents for large volumes of biomedical literature, using the MMTV-like virus and breast cancer association as a focused case study. We do not claim to have implemented or validated the full AI co-scientist pipeline end to end. Moreover, our evaluation is limited to a single microbe–cancer pair within the biomedical domain, whereas a general AI co-scientist is expected to operate across diverse scientific domains and problem types. While we hypothesize that the proposed filtering approach may generalize to other microbe–cancer associations, this remains to be empirically validated.
It is also important to note that the reasoning capabilities of SLMs and LLMs are shaped by the nature and distribution of their pre-training data. As a result, well-studied microbe–cancer pairs may be easier for such models to recognize and reason about, whereas rarer or more niche associations may present greater difficulty. Consequently, model performance may vary across different microbe–cancer pairs, and the results observed in this study may not directly transfer to all MCPs. Future work should extend this framework to additional biomedical relationships and to other scientific domains to assess the broader viability of the AI co-scientist paradigm.

While expert-driven consensus was used to establish ground truth labels, our analysis revealed inevitable ambiguities at the boundaries between classification categories. The strict adherence to nuanced criteria by some models (particularly GPT-5) exposed cases where expert judgment involves subtle interpretation—such as distinguishing whether a paper mentioning MMTV promoter sequences in transgenic mouse models constitutes evidence "mentioning MMTV" as a microbe versus merely using a genetic element from the virus. Such edge cases highlight a fundamental challenge: mapping complex scientific abstracts onto discrete categorical labels necessarily involves a degree of interpretive flexibility that even domain experts may resolve differently. For instance, GPT-5's classification of papers providing negative evidence (no association detected) as "somewhat relevant" rather than "relevant" reflects a stricter interpretation of "causal evidence" than our expert annotations applied, yet represents a defensible reading of the criteria. While we believe our expert consensus labels represent appropriate scientific judgment for this task, we acknowledge that alternative reasonable interpretations exist for a subset of borderline cases. This inherent ambiguity in nuanced classification tasks means that perfect inter-annotator agreement—whether between humans or between humans and models—may be unattainable, and models exhibiting high disagreement with our labels are not necessarily performing poorly but may instead be applying the criteria with different interpretive priorities.

The total analysis times reported for BFS-RS with SLMs in Section~\ref{sec:models_times} do not reflect the most computationally efficient configuration of the method. Specifically, the hyperparameter tuning process required between 100 and 200 hours because BFS-RS was executed using a single thread. This conservative choice was made intentionally to avoid potential race conditions and to ensure deterministic behavior, as initial experiments revealed a degree of stochasticity in the BFS-RS results. While this setting ensured reproducibility, it significantly increased the overall runtime. We strongly believe that executing BFS-RS with multi-threading and on modern GPU hardware would drastically reduce the total tuning time and would not require on the order of 100–200 hours.

\section{Conclusion}\label{sec:conclusion}

This study demonstrates that modern Small Language Models (SLMs, $\leq$ 8B parameters) can perform nuanced biomedical literature classification with accuracy and reliability that approach state-of-the-art frontier Large Language Models (LLMs). Using the challenging case of HMTV/MMTV-like virus and breast cancer, we show that models such as Llama 3 and Qwen2.5 achieve strong zero-shot performance outperforming Gemini 3 Pro Preview, GPT-5 and Meerkat, and that with simple few-shot and programmatic prompt optimization strategies, they can match or closely trail Gemini 2.5 Pro on key filtering metrics. In addition, this research contributes a high-quality, expert-validated benchmark dataset for evaluating classification capabilities on complex biomedical reasoning tasks.

Our results reveal that performance gains depend strongly on both model architecture and prompting strategy. Qwen2.5 consistently benefits from explicit reasoning and bootstrapped few-shot optimization (BFS-RS), while Llama 3 responds most strongly to principled demonstration selection such as fair and central sampling. Perturbation-based interpretability shows that SLM decisions are often grounded in valid biomedical cues, though sensitivity to spurious lexical artifacts highlights the importance of transparency and careful validation in high-stakes scientific workflows.

Taken together, these findings establish SLMs as viable, cost-efficient front-end filters within AI co-scientist pipelines, enabling scalable and reproducible biomedical literature triage. Beyond this specific case study, the broader implication of our findings is their relevance to scalable discovery of novel Microbe–Cancer Pairs (MCP). Efficient and reliable automated screening is a critical first step in identifying previously unrecognized microbial contributions to cancer—discoveries that carry immense potential for life- and cost-saving interventions through prevention, vaccination, and targeted public health strategies. By enabling reproducible filtering across many candidate MCPs, SLM-driven pipelines can potentially provide a practical foundation for accelerating systematic discovery across the biomedical literature.

\section{Future Work}\label{sec:futurework}

Our findings open several promising avenues for future research spanning prompt optimization, model specialization, interpretability, domain generalization, and real-world deployment.

\subsubsection*{\textbf{Advancing prompt optimization techniques:}} The substantial performance gains achieved by Qwen2.5 under BFS-RS demonstrate that programmatic prompt optimization can unlock latent capabilities in SLMs.Future work should systematically explore alternative optimizers beyond BFS-RS, including DSPy’s MIPRO \cite{opsahl2024optimizing} and advanced optimizers in frameworks such as AdalFlow \cite{adalflow}. Moreover, our study employed a relatively simple DSPy program based on a single Chain-of-Thought module. Investigating more complex multi-stage pipelines—incorporating retrieval, multi-step reasoning, or ensemble strategies may yield further gains in performance.

\subsubsection*{\textbf{Addressing classification ambiguity through expert panels:}}
The unexpectedly poor performance of GPT-5's reasoning modes compared to smaller models highlights the inherent subjectivity at classification boundaries, where even experts may interpret subtle cues differently (as noted in Section~\ref{sec:limitations}). Such ambiguities suggest that some disagreements between models and ground truth labels reflect defensible alternative interpretations rather than model failure. Future work could explore structured approaches to resolving these ambiguities, such as leveraging panels of multiple experts to reach consensus on borderline cases or incorporating uncertainty-aware labeling schemes. Such approaches may provide more robust ground truth for training and evaluating models, particularly when dealing with nuanced scientific criteria.

\subsubsection*{\textbf{Improving domain-specialized models:}} Meerkat’s poor overall performance despite medical fine-tuning suggests that domain specialization alone does not guarantee robust reasoning. Future work should examine whether applying in-context learning (ICL) and prompt optimization frameworks to domain-specialized models can better leverage their medical knowledge. Comparative studies assessing whether further fine-tuning general-purpose SLMs on domain-specific corpora outperforms pre-specialized models would clarify the relative value of adaptation versus general reasoning ability.

\subsubsection*{\textbf{Expanding model coverage and computational scale:}}
Due to computational constraints, this study only focused on Llama 3 and Qwen2.5. Future work should evaluate additional SLMs (e.g., Mistral, DeepSeek-R1, Qwen3) to assess whether observed trends in ICL responsiveness, optimization sensitivity, and perturbation robustness generalize across architectures. Additionally, scaling BFS-RS to larger candidate pools beyond our current $N = 150$ could determine whether more exhaustive search yields diminishing or meaningful returns.

\subsubsection*{\textbf{Deepening interpretability analysis:}} Our perturbation-based analysis revealed sensitivity to spurious lexical features, but a fuller understanding of SLM decision mechanisms requires complementary interpretability techniques. Future work should integrate attention analysis, gradient-based attribution, and counterfactual explanations to triangulate model behavior. Determining whether observed brittleness is intrinsic to current architectures or reducible through interventions such as adversarial training or explanation-regularized learning will be critical for high-stakes scientific deployment.

\subsubsection*{\textbf{Generalization to broader domains and novel Microbe-Cancer Pair (MCP) discovery:}} While this study focused on the HMTV/MMTV-like virus and breast cancer, future research should assess generalization to other highly structured, criteria-driven domains such as drug–target interaction discovery, climate attribution studies, and materials property prediction. Most importantly, this work represents an initial step toward large-scale mining of the biomedical literature for identifying novel, high-impact microbe–cancer pairs (MCPs). Because new discoveries of microbial facilitation of cancers—especially those linked to high mortality and disability—offer immense life- and cost-saving potential through prevention, vaccination, and targeted public health strategies, future studies could build on this pipeline to systematically explore and identify many candidate MCPs across diverse contexts.

\subsubsection*{\textbf{Model Ensembling:}}
 Future work should explore whether ensembling multiple SLMs can further improve robustness and accuracy. Simple majority voting schemes could serve as a baseline, while more advanced stacking approaches—where a meta-model integrates outputs from multiple base models—may better exploit model diversity.

\subsubsection*{\textbf{Scaling to operational deployment:}} Finally, future work should investigate deployment considerations beyond classification accuracy. This includes developing effective human–AI collaboration strategies (e.g., active learning for expert-in-the-loop review), reliable calibration methods for prediction confidence, and prospective evaluations in real systematic review workflows to measure downstream impact on research synthesis quality and efficiency.

By pursuing these directions, future research can advance toward reliable, interpretable, and practically deployable AI co-scientist systems for supporting large-scale scientific discovery.

\section*{Acknowledgements}
We thank Caryn Mcnamara for general research and administrative support. This work was supported by the Infectious Diseases and Oncology Research Institute (IDORI). Kaela Kokkas, Mohammad Zaid Moonsamy and Muhammed Muaaz Dawood also acknowledge Wits Health Consortium (Pty) Limited (WHC) for MSc and PhD fellowship support. Opinions expressed and conclusions arrived at are those of the author and are not necessarily to be attributed to IDORI.

\bibliographystyle{ACM-Reference-Format}
\bibliography{references}

\appendix
\label{sec:appendix}

\section*{Appendices}

\section{Computational Resources and Hardware Configuration}\label{app:hardware_timing}

\subsection{Hardware Specifications for Timing Benchmarks (Zero-Shot and Final Evaluation)}
The timing and performance benchmarks used for initial model selection (Section~\ref{sec:models_times})—including per-paper processing speed for Llama 3 and Qwen2.5—were conducted on a single laptop to ensure consistency and transparent reporting of feasibility constraints. The hardware specifications are as follows:
\begin{itemize}
    \item \textbf{CPU}: AMD Ryzen 7 5800H
    \item \textbf{System RAM}: 16 GB
    \item \textbf{GPU}: NVIDIA GeForce RTX 3060 Laptop GPU
    \item \textbf{VRAM}: 6 GB GDDR6
\end{itemize}

\subsection{Hardware Specifications for BFS-RS Optimization Experiments}
The time-intensive hyperparameter tuning and optimization runs for the Bootstrap Few-Shot with Random Search (BFS-RS) strategy (total analysis time of 100 to 200 hours, as noted in Section~\ref{sec:models_times}) were executed on dedicated nodes within a high-performance computing cluster. The specifications of the utilized nodes used are summarised below:

\begin{itemize}
    \item \textbf{System}: High-Performance Computing Cluster (bigbatch queue)
    \item \textbf{Node Specifications} (per node):
    \begin{itemize}
        \item \textbf{CPU}: Intel Core i9-10940X (14 cores)
        \item \textbf{System RAM}: 128 GB
        \item \textbf{GPU}: NVIDIA RTX 3090
        \item \textbf{VRAM}: 24 GB
    \end{itemize}
    \item \textbf{Environment Constraint}: All BFS-RS optimization runs were restricted to a single thread ($\text{num\_threads}=1$) to ensure rigorous and reproducible results and eliminate potential race conditions, despite the higher computational resources available.
\end{itemize}

This configuration enabled efficient management of the extensive hyperparameter search across multiple candidate configurations and random seeds.

\section{Formal Logic Formulation of the  Prompt}\label{sec:logic}

In this paper we have explored the power of SLMs to use nuanced reasoning to classify papers. Although we have used a prompt in section \ref{sec:prompts} targeting  a specific microbe-cancer pair, the prompt framework is general. To illustrate this we rewrite the specific prompt in general form here using more formal logic:

\textbf{Definitions:}
\begin{itemize}
    \item 
Let $R$ be a document (article) to be classified.
    \item 
Let $M$ denote the target microbe (e.g. HMTV/MMTV-like virus).
    \item 
Let $\mathcal{F}(M)$ denote the set of microbes belonging to the same
species/family as $M$.
    \item 
Let $C$ denote the specific cancer of interest (e.g. breast cancer).
    \item 
Let $\mathcal{C_{\textit{All}}}$ denote the set of all cancers.
\end{itemize}
We then define the following predicates:
\begin{align*}
R_1(R) &:=
\text{$R$ gives empirical causal evidence about whether } 
M \text{ or $\mathcal{F}(M)$ cause $C$ or any cancer in $\mathcal{C_{\textit{All}}}$}. \\
R_2(R) &:=
\text{$R$ gives a theoretical causal model for how } M \text{ could cause $C$}. \\
S_1(R) &:=
\text{$R$ mentions both } M \text{ and $C$ but provides no causal evidence or mechanism}. \\
S_2(R) &:=
\text{$R$ gives causal evidence about other microbes causing $C$,}  \text{and these findings can plausibly generalize to  $M$.}
\end{align*}


\textbf{Class Definitions:}

A document $R$ is assigned to the Relevant, Somewhat Relevant, or Irrelevant class according to the following mutually exclusive Boolean predicates:: 
\[
\mathrm{Relevant}(R) \iff R1(R) \lor R2(R).
\]
\[
\mathrm{SomewhatRelevant}(R) 
\iff \neg \mathrm{Relevant}(R) \land (S1(R) \lor S2(R)).
\]
\[
\mathrm{Irrelevant}(R) 
\iff \neg R1(R) \land \neg R2(R) \land \neg S1(R) \land \neg S2(R).
\]

Here $\iff, \land, \lor$ and $\neg$ represent the logical operators "If and only if" (logically equivalent), \textbf{AND}, \textbf{OR} and \textbf{NOT} respectively.  

\section{BFS-RS Experimentation Details}\label{app:bfs-rs-additional}

\subsection{DSPy Signature and Prompt Initialization}
In our implementation of the BFS-RS (Bootstrap Few-shot with Random Search) optimizer and the zero-shot DSPy implementations, we utilized the DSPy framework to manage programmatic prompting and modular reasoning. The DSPy framework employs \textbf{Signatures} to define the input–output schema of a task and supports multiple composable reasoning modules such as \texttt{dspy.Predict} and \texttt{dspy.CoT}. 

For our study, we used the Chain-of-Thought (CoT) module to enable step-by-step reasoning during prediction, and a custom \texttt{QA} Signature to specify the task prompt and output format. These components are shown below to ensure reproducibility and transparency in our experimental setup.



\begin{lstlisting}[frame=single, framesep=5pt]{python}
class CoT(dspy.Module):
    def __init__(self):
        super().__init__()
        self.prog = dspy.ChainOfThought(QA)   # dspy.Predict(QA)

    def forward(self, title, abstract):
        return self.prog(title=title, abstract=abstract)
\end{lstlisting}



\begin{lstlisting}[frame=single, framesep=5pt, breaklines]{python}
class QA(dspy.Signature):
    "How relevant is this article in determining whether it is plausible that HMTV/MMTV-like virus can cause breast cancer in humans(relevant, somewhat relevant, or irrelevant)? A relevant article would fulfil either of the following criteria: 1) gives causal evidence for whether or not HMTV/MMTV-like virus or its species/family of microbes cause breast cancer or other cancers either in humans or animal models, or 2) contains a theoretical model for how HMTV/MMTV-like virus might cause breast cancer, irrespective of whether there is experimental evidence. A somewhat relevant article would fulfil either of the following criteria: 1) mentions both HMTV/MMTV-like virus and breast cancer, but does not give any causal arguments or evidence for HMTV/MMTV-like virus causing breast cancer, or 2) contains evidence examining whether other microbes cause breast cancer, if the findings can plausibly generalise to HMTV/MMTV-like virus and breast cancer. An irrelevant article would be a paper that does not fulfil the criteria for the relevant or somewhat relevant classifications. FORMAT: In your answer, use only RELEVANT, SOMEWHAT RELEVANT, IRRELEVANT or NO CLASSIFICATION as your response."

    title = dspy.InputField(desc="Title of the article")
    abstract = dspy.InputField(desc="Abstract of the article")
    label = dspy.OutputField(desc="use only RELEVANT, SOMEWHAT RELEVANT, IRRELEVANT or NO CLASSIFICATION as your response")
\end{lstlisting}

\subsection{Additional details on Bootstrap Few Shot with Random Search (BFS-RS)}

The Bootstrap Few Shot with Random Search (BFS-RS) optimizer (described conceptually in Section~\ref{sec:few-shot-methods}) identifies high-quality few-shot demonstrations to include in the prompt. This optimizer requires a training set of input-output pairs and two key parameters: $max\_labeled\_demos$ (the total number of demonstrations $k$ in the final prompt) and $max\_bootstrapped\_demos$ (the number of bootstrapped demonstrations $b$ in the final prompt, where $b \leq k$). The Random Search (RS) component addresses order-sensitivity by running the bootstrapping process multiple times with shuffled training data, evaluating each candidate set on a validation set to select the best-performing configuration.

Two important details merit clarification. First, when the teacher model generates bootstrapped demonstrations, it uses few-shot examples drawn from the training set as part of its own prompt, with the $max\_labeled\_demos$ parameter controlling how many such examples are included; however, the teacher always excludes the specific sample for which it is currently generating a reasoning trace to avoid data leakage. Second, the final optimized prompt contains between one and $B$ bootstrapped demonstrations (formally, $b \in \{1, ..., B\}$), with the exact number determined by the optimization process—specifically, by how many successful bootstrapped demonstrations the teacher model produces during its iteration through the training set. The resulting prompt therefore includes $k$ total samples, comprising $b$ bootstrapped demonstrations and $(k-b)$ additional labeled examples if $b < k$.

\subsection{BFS-RS Configuration and Hyperparamater tuning} 

To ensure full reproducibility, all runs were executed in single-threaded mode by setting $num\_threads$ parameter to 1, thereby eliminating potential inconsistencies and race conditions. The $max\_rounds$ and $max\_errors$ parameters were set to 3, allowing sufficient iterations to identify the optimal number of bootstrapped demonstrations while handling ocassional failures. The optimizer employed the exact match (EM) \cite{khattab2024dspy} metric for demonstration evaluation, which provides a direct binary measure well suited to our discrete class labels.

We set $max\_labeled\_demos$ and $max\_bootstrapped\_demos$ to identical values to simplify parameter tuning. Since $max\_bootstrapped\_demos$ can range from 1 to $b$ (where $b$ represents the number of bootstrapped demonstrations), constraining it to equal the $max\_labeled\_demos$ value $k$ ensures that $b \leq k$. Consequently, we tuned both parameters jointly as a single hyperparameter.  We used the external validation set exclusively for hyperparameter tuning, specifically optimizing $max\_labeled\_demos / max\_bootstrapped\_demos$ and $num\_candidates$.

Table~\ref{tab:val_set_one} reports the initial hyperparameter search across varying $N$ and $k$ values, revealing that $N = 100$ produced the most stable and accurate results for both models. Table~\ref{tab:val_set_two} presents the refined tuning phase, in which adjacent $k$ values around the best configuration were evaluated to determine the final optimal setup. Results indicate that the best-performing configurations corresponded to $N = 100$, with optimal $k$ values of 3 for Llama 3 and 4 for Qwen2.5.

\begin{table}[htbp]
\centering
\caption{Three-class classification results for bootstrap few shot with random search (BFS-RS) on the external validation set (30 samples). The standard deviations are obtained across 10 different random seeds (42-52). $N$ denotes the number of candidates explored, while $k$ denotes the number of demonstrations (samples) appended to the prompt. We tested $k \in \{2, 5, 8\}$ for each $N \in \{13, 30, 50, 75, 100, 150\}$. The highest values in each column are shown in bold. The best $N$ and $k$ value, along with the corresponding model names, are also bolded. The optimal value of $N$ was 100, while the optimal $k$ was 2 for Llama 3 and 5 for Qwen2.5.}
\label{tab:val_set_one}
\resizebox{\textwidth}{!}{%
\begin{tabular}{lccccccccc}
\hline
\textbf{Model} & \textbf{N} & \textbf{k} & \textbf{Accuracy} & \textbf{Precision} & \textbf{Sensitivity} & \textbf{Specificity} & \textbf{F0.2-Score} & \textbf{F1-Score} & \textbf{F5-Score} \\
\toprule
Llama 3 & 13 & 2 & 0.70 $\pm$ 0.10 & 0.59 $\pm$ 0.18 & 0.52 $\pm$ 0.15 & 0.88 $\pm$ 0.03 & 0.58 $\pm$ 0.17 & 0.52 $\pm$ 0.15 & 0.52 $\pm$ 0.15 \\
Llama 3 & 13 & 5 & 0.73 $\pm$ 0.04 & 0.59 $\pm$ 0.14 & 0.58 $\pm$ 0.09 & 0.86 $\pm$ 0.03 & 0.58 $\pm$ 0.13 & 0.56 $\pm$ 0.10 & 0.58 $\pm$ 0.09 \\
Llama 3 & 13 & 8 & 0.64 $\pm$ 0.06 & 0.53 $\pm$ 0.08 & 0.53 $\pm$ 0.06 & 0.81 $\pm$ 0.03 & 0.53 $\pm$ 0.08 & 0.51 $\pm$ 0.06 & 0.53 $\pm$ 0.06 \\
\cmidrule{3-10}
Llama 3 & 30 & 2 & 0.69 $\pm$ 0.10 & 0.55 $\pm$ 0.15 & 0.51 $\pm$ 0.16 & 0.89 $\pm$ 0.02 & 0.55 $\pm$ 0.15 & 0.52 $\pm$ 0.15 & 0.51 $\pm$ 0.16 \\
Llama 3 & 30 & 5 & 0.71 $\pm$ 0.03 & 0.52 $\pm$ 0.09 & 0.56 $\pm$ 0.07 & 0.85 $\pm$ 0.02 & 0.51 $\pm$ 0.08 & 0.52 $\pm$ 0.07 & 0.55 $\pm$ 0.07 \\
Llama 3 & 30 & 8 & 0.66 $\pm$ 0.06 & 0.51 $\pm$ 0.08 & 0.54 $\pm$ 0.06 & 0.82 $\pm$ 0.03 & 0.51 $\pm$ 0.08 & 0.51 $\pm$ 0.06 & 0.54 $\pm$ 0.06 \\
\cmidrule{3-10}
Llama 3 & 50 & 2 & 0.67 $\pm$ 0.10 & 0.58 $\pm$ 0.16 & 0.51 $\pm$ 0.14 & 0.87 $\pm$ 0.04 & 0.57 $\pm$ 0.15 & 0.51 $\pm$ 0.13 & 0.51 $\pm$ 0.14 \\
Llama 3 & 50 & 5 & 0.73 $\pm$ 0.05 & 0.57 $\pm$ 0.14 & 0.59 $\pm$ 0.07 & 0.85 $\pm$ 0.03 & 0.57 $\pm$ 0.12 & 0.56 $\pm$ 0.08 & 0.58 $\pm$ 0.07 \\
Llama 3 & 50 & 8 & 0.72 $\pm$ 0.05 & 0.51 $\pm$ 0.11 & 0.57 $\pm$ 0.08 & 0.85 $\pm$ 0.02 & 0.51 $\pm$ 0.10 & 0.52 $\pm$ 0.07 & 0.56 $\pm$ 0.07 \\
\cmidrule{3-10}
Llama 3 & 75 & 2 & 0.68 $\pm$ 0.12 & 0.52 $\pm$ 0.11 & 0.48 $\pm$ 0.12 & 0.88 $\pm$ 0.04 & 0.52 $\pm$ 0.10 & 0.48 $\pm$ 0.10 & 0.47 $\pm$ 0.12 \\
Llama 3 & 75 & 5 & 0.72 $\pm$ 0.03 & 0.56 $\pm$ 0.11 & 0.57 $\pm$ 0.06 & 0.85 $\pm$ 0.01 & 0.55 $\pm$ 0.10 & 0.54 $\pm$ 0.05 & 0.57 $\pm$ 0.06 \\
Llama 3 & 75 & 8 & 0.71 $\pm$ 0.05 & 0.51 $\pm$ 0.11 & 0.56 $\pm$ 0.08 & 0.84 $\pm$ 0.02 & 0.51 $\pm$ 0.10 & 0.52 $\pm$ 0.07 & 0.56 $\pm$ 0.07 \\
\cmidrule{3-10}
\textbf{Llama 3} & \textbf{100} & \textbf{2} & \textbf{0.74 $\pm$ 0.05} & \textbf{0.62 $\pm$ 0.11} & \textbf{0.57 $\pm$ 0.09} & \textbf{0.88 $\pm$ 0.03} & \textbf{0.61 $\pm$ 0.10} & \textbf{0.57 $\pm$ 0.09} & \textbf{0.57 $\pm$ 0.09} \\
Llama 3 & 100 & 5 & 0.72 $\pm$ 0.03 & 0.53 $\pm$ 0.07 & 0.58 $\pm$ 0.02 & 0.84 $\pm$ 0.02 & 0.53 $\pm$ 0.06 & 0.54 $\pm$ 0.03 & 0.57 $\pm$ 0.02 \\
Llama 3 & 100 & 8 & 0.71 $\pm$ 0.05 & 0.51 $\pm$ 0.11 & 0.56 $\pm$ 0.08 & 0.84 $\pm$ 0.02 & 0.51 $\pm$ 0.10 & 0.52 $\pm$ 0.07 & 0.56 $\pm$ 0.07 \\
\cmidrule{3-10}
Llama 3 & 150 & 2 & 0.71 $\pm$ 0.05 & 0.58 $\pm$ 0.08 & 0.55 $\pm$ 0.08 & 0.88 $\pm$ 0.03 & 0.58 $\pm$ 0.08 & 0.55 $\pm$ 0.08 & 0.55 $\pm$ 0.08 \\
Llama 3 & 150 & 5 & 0.73 $\pm$ 0.03 & 0.57 $\pm$ 0.11 & 0.59 $\pm$ 0.06 & 0.84 $\pm$ 0.02 & 0.56 $\pm$ 0.10 & 0.56 $\pm$ 0.04 & 0.59 $\pm$ 0.03 \\
Llama 3 & 150 & 8 & 0.73 $\pm$ 0.06 & 0.50 $\pm$ 0.08 & 0.58 $\pm$ 0.08 & 0.86 $\pm$ 0.03 & 0.50 $\pm$ 0.07 & 0.53 $\pm$ 0.08 & 0.57 $\pm$ 0.08 \\
\midrule
Qwen2.5 & 13 & 2 & 0.69 $\pm$ 0.04 & 0.68 $\pm$ 0.03 & 0.63 $\pm$ 0.06 & 0.85 $\pm$ 0.02 & 0.67 $\pm$ 0.04 & 0.64 $\pm$ 0.05 & 0.63 $\pm$ 0.06 \\
Qwen2.5 & 13 & 5 & 0.74 $\pm$ 0.09 & 0.72 $\pm$ 0.08 & 0.69 $\pm$ 0.11 & 0.87 $\pm$ 0.04 & 0.72 $\pm$ 0.08 & 0.69 $\pm$ 0.10 & 0.69 $\pm$ 0.11 \\
Qwen2.5 & 13 & 8 & 0.67 $\pm$ 0.00 & 0.66 $\pm$ 0.00 & 0.61 $\pm$ 0.00 & 0.84 $\pm$ 0.00 & 0.66 $\pm$ 0.00 & 0.61 $\pm$ 0.00 & 0.61 $\pm$ 0.00 \\
\cmidrule{3-10}
Qwen2.5 & 30 & 2 & 0.70 $\pm$ 0.05 & 0.69 $\pm$ 0.04 & 0.65 $\pm$ 0.06 & 0.86 $\pm$ 0.03 & 0.68 $\pm$ 0.04 & 0.65 $\pm$ 0.05 & 0.65 $\pm$ 0.06 \\
Qwen2.5 & 30 & 5 & 0.80 $\pm$ 0.07 & 0.78 $\pm$ 0.07 & 0.77 $\pm$ 0.09 & 0.90 $\pm$ 0.04 & 0.77 $\pm$ 0.07 & 0.75 $\pm$ 0.08 & 0.77 $\pm$ 0.09 \\
Qwen2.5 & 30 & 8 & 0.67 $\pm$ 0.00 & 0.66 $\pm$ 0.00 & 0.61 $\pm$ 0.00 & 0.84 $\pm$ 0.00 & 0.66 $\pm$ 0.00 & 0.61 $\pm$ 0.00 & 0.61 $\pm$ 0.00 \\
\cmidrule{3-10}
Qwen2.5 & 50 & 2 & 0.72 $\pm$ 0.05 & 0.70 $\pm$ 0.04 & 0.68 $\pm$ 0.07 & 0.87 $\pm$ 0.03 & 0.70 $\pm$ 0.04 & 0.67 $\pm$ 0.06 & 0.68 $\pm$ 0.07 \\
Qwen2.5 & 50 & 5 & 0.80 $\pm$ 0.09 & 0.78 $\pm$ 0.09 & 0.77 $\pm$ 0.12 & 0.91 $\pm$ 0.04 & 0.77 $\pm$ 0.10 & 0.75 $\pm$ 0.11 & 0.77 $\pm$ 0.12 \\
Qwen2.5 & 50 & 8 & 0.67 $\pm$ 0.00 & 0.66 $\pm$ 0.00 & 0.61 $\pm$ 0.00 & 0.84 $\pm$ 0.00 & 0.66 $\pm$ 0.00 & 0.61 $\pm$ 0.00 & 0.61 $\pm$ 0.00 \\
\cmidrule{3-10}
Qwen2.5 & 75 & 2 & 0.79 $\pm$ 0.06 & 0.76 $\pm$ 0.05 & 0.76 $\pm$ 0.08 & 0.90 $\pm$ 0.03 & 0.76 $\pm$ 0.06 & 0.75 $\pm$ 0.07 & 0.76 $\pm$ 0.07 \\
Qwen2.5 & 75 & 5 & 0.80 $\pm$ 0.07 & 0.78 $\pm$ 0.07 & 0.78 $\pm$ 0.09 & 0.91 $\pm$ 0.04 & 0.78 $\pm$ 0.07 & 0.76 $\pm$ 0.08 & 0.77 $\pm$ 0.09 \\
Qwen2.5 & 75 & 8 & 0.67 $\pm$ 0.00 & 0.66 $\pm$ 0.00 & 0.61 $\pm$ 0.00 & 0.84 $\pm$ 0.00 & 0.66 $\pm$ 0.00 & 0.61 $\pm$ 0.00 & 0.61 $\pm$ 0.00 \\
\cmidrule{3-10}
Qwen2.5 & 100 & 2 & 0.80 $\pm$ 0.03 & 0.77 $\pm$ 0.03 & 0.79 $\pm$ 0.03 & 0.91 $\pm$ 0.01 & 0.77 $\pm$ 0.03 & 0.76 $\pm$ 0.03 & 0.78 $\pm$ 0.03 \\
\textbf{Qwen2.5} & \textbf{100} & \textbf{5} & \textbf{0.83 $\pm$ 0.07} & \textbf{0.80 $\pm$ 0.08} & \textbf{0.80 $\pm$ 0.10} & \textbf{0.92 $\pm$ 0.03} & \textbf{0.80 $\pm$ 0.08} & \textbf{0.78 $\pm$ 0.09} & \textbf{0.80 $\pm$ 0.10} \\
Qwen2.5 & 100 & 8 & 0.67 $\pm$ 0.00 & 0.66 $\pm$ 0.00 & 0.61 $\pm$ 0.00 & 0.84 $\pm$ 0.00 & 0.66 $\pm$ 0.00 & 0.61 $\pm$ 0.00 & 0.61 $\pm$ 0.00 \\
\cmidrule{3-10}
Qwen2.5 & 150 & 2 & 0.80 $\pm$ 0.04 & 0.77 $\pm$ 0.04 & 0.78 $\pm$ 0.05 & 0.91 $\pm$ 0.02 & 0.77 $\pm$ 0.04 & 0.76 $\pm$ 0.05 & 0.78 $\pm$ 0.05 \\
Qwen2.5 & 150 & 5 & 0.82 $\pm$ 0.07 & 0.79 $\pm$ 0.08 & 0.79 $\pm$ 0.09 & 0.92 $\pm$ 0.03 & 0.79 $\pm$ 0.08 & 0.78 $\pm$ 0.09 & 0.79 $\pm$ 0.09 \\
Qwen2.5 & 150 & 8 & 0.68 $\pm$ 0.04 & 0.68 $\pm$ 0.07 & 0.61 $\pm$ 0.02 & 0.84 $\pm$ 0.01 & 0.67 $\pm$ 0.05 & 0.62 $\pm$ 0.02 & 0.61 $\pm$ 0.02 \\
\bottomrule
\end{tabular}%
}
\end{table}

\begin{table}[htbp]
\centering
\caption{Three-class classification results for bootstrap few shot with random search (BFS-RS) on the external validation set of size 30 samples for adaptive hyperparameter tuning. Following the identification of the optimal $N$ (number of candidates explored) value in Table~\ref{tab:val_set_one}, adjacent $k$ (number of demonstrations appended to the prompt) values are evaluated to refine the configuration. The highest values in each column are shown in bold. The best $N$ and $k$ value, along with the corresponding model names, are also bolded. The optimal value of $k$ was 3 for Llama 3 and 4 for Qwen2.5.}
\label{tab:val_set_two}
\resizebox{\textwidth}{!}{%
\begin{tabular}{lccccccccc}
\toprule
\textbf{Model} & \textbf{N} & \textbf{k} & \textbf{Accuracy} & \textbf{Precision} & \textbf{Sensitivity} & \textbf{Specificity} & \textbf{F0.2-Score} & \textbf{F1-Score} & \textbf{F5-Score} \\
\midrule

Llama 3 & 100 & 1 & 0.73 $\pm$ 0.02 & 0.55 $\pm$ 0.05 & 0.54 $\pm$ 0.03 & 0.90 $\pm$ 0.03 & 0.55 $\pm$ 0.04 & 0.54 $\pm$ 0.03 & 0.54 $\pm$ 0.03 \\

Llama 3 & 100 & 2 & 0.74 $\pm$ 0.05 & 0.62 $\pm$ 0.11 & 0.57 $\pm$ 0.09 & 0.88 $\pm$ 0.03 & 0.61 $\pm$ 0.10 & 0.57 $\pm$ 0.09 & 0.57 $\pm$ 0.09 \\

\textbf{Llama 3} & \textbf{100} & \textbf{3} & \textbf{0.76 $\pm$ 0.04} & \textbf{0.61 $\pm$ 0.13} & \textbf{0.61 $\pm$ 0.08} & \textbf{0.88 $\pm$ 0.02} & \textbf{0.61 $\pm$ 0.12} & \textbf{0.60 $\pm$ 0.08} & \textbf{0.61 $\pm$ 0.08} \\

Llama 3 & 100 & 4 & 0.73 $\pm$ 0.09 & 0.59 $\pm$ 0.19 & 0.59 $\pm$ 0.14 & 0.87 $\pm$ 0.03 & 0.59 $\pm$ 0.18 & 0.57 $\pm$ 0.15 & 0.58 $\pm$ 0.14 \\

Llama 3 & 100 & 5 & 0.72 $\pm$ 0.03 & 0.53 $\pm$ 0.07 & 0.58 $\pm$ 0.02 & 0.84 $\pm$ 0.02 & 0.53 $\pm$ 0.06 & 0.54 $\pm$ 0.03 & 0.57 $\pm$ 0.02 \\

Llama 3 & 100 & 8 & 0.71 $\pm$ 0.05 & 0.51 $\pm$ 0.11 & 0.56 $\pm$ 0.08 & 0.84 $\pm$ 0.02 & 0.51 $\pm$ 0.10 & 0.52 $\pm$ 0.07 & 0.56 $\pm$ 0.07 \\

\midrule

Qwen2.5 & 100 & 2 & 0.80 $\pm$ 0.03 & 0.77 $\pm$ 0.03 & 0.79 $\pm$ 0.03 & 0.91 $\pm$ 0.01 & 0.77 $\pm$ 0.03 & 0.76 $\pm$ 0.03 & 0.78 $\pm$ 0.03 \\

Qwen2.5 & 100 & 3 & 0.79 $\pm$ 0.09 & 0.76 $\pm$ 0.08 & 0.75 $\pm$ 0.10 & 0.90 $\pm$ 0.04 & 0.76 $\pm$ 0.08 & 0.74 $\pm$ 0.10 & 0.75 $\pm$ 0.10 \\

\textbf{Qwen2.5} & \textbf{100} & \textbf{4} & \textbf{0.85 $\pm$ 0.08} & \textbf{0.83 $\pm$ 0.09} & \textbf{0.82 $\pm$ 0.09} & \textbf{0.93 $\pm$ 0.04} & \textbf{0.83 $\pm$ 0.09} & \textbf{0.82 $\pm$ 0.09} & \textbf{0.82 $\pm$ 0.09} \\

Qwen2.5 & 100 & 5 & 0.83 $\pm$ 0.07 & 0.80 $\pm$ 0.08 & 0.80 $\pm$ 0.10 & 0.92 $\pm$ 0.03 & 0.80 $\pm$ 0.08 & 0.78 $\pm$ 0.09 & 0.80 $\pm$ 0.10 \\

Qwen2.5 & 100 & 6 & 0.83 $\pm$ 0.10 & 0.81 $\pm$ 0.08 & 0.80 $\pm$ 0.13 & 0.92 $\pm$ 0.05 & 0.80 $\pm$ 0.09 & 0.78 $\pm$ 0.13 & 0.80 $\pm$ 0.13 \\

Qwen2.5 & 100 & 7 & 0.83 $\pm$ 0.08 & 0.84 $\pm$ 0.09 & 0.78 $\pm$ 0.08 & 0.91 $\pm$ 0.03 & 0.83 $\pm$ 0.09 & 0.79 $\pm$ 0.08 & 0.78 $\pm$ 0.08 \\

Qwen2.5 & 100 & 8 & 0.67 $\pm$ 0.00 & 0.66 $\pm$ 0.00 & 0.61 $\pm$ 0.00 & 0.84 $\pm$ 0.00 & 0.66 $\pm$ 0.00 & 0.61 $\pm$ 0.00 & 0.61 $\pm$ 0.00 \\
\bottomrule
\end{tabular}%
}
\end{table}

\section{Additional Perturbation Analysis}\label{app:perturbation}
To assess whether the patterns observed with Qwen2.5 generalize to Llama 3, we conducted the perturbation analysis on Llama 3. Across both models, we observe the same underlying characteristics: accurate identification of key scientific terms coexisting with sensitivity to spurious textual features.

\begin{figure}[ht]
    \centering
    \includegraphics[width=1\linewidth]{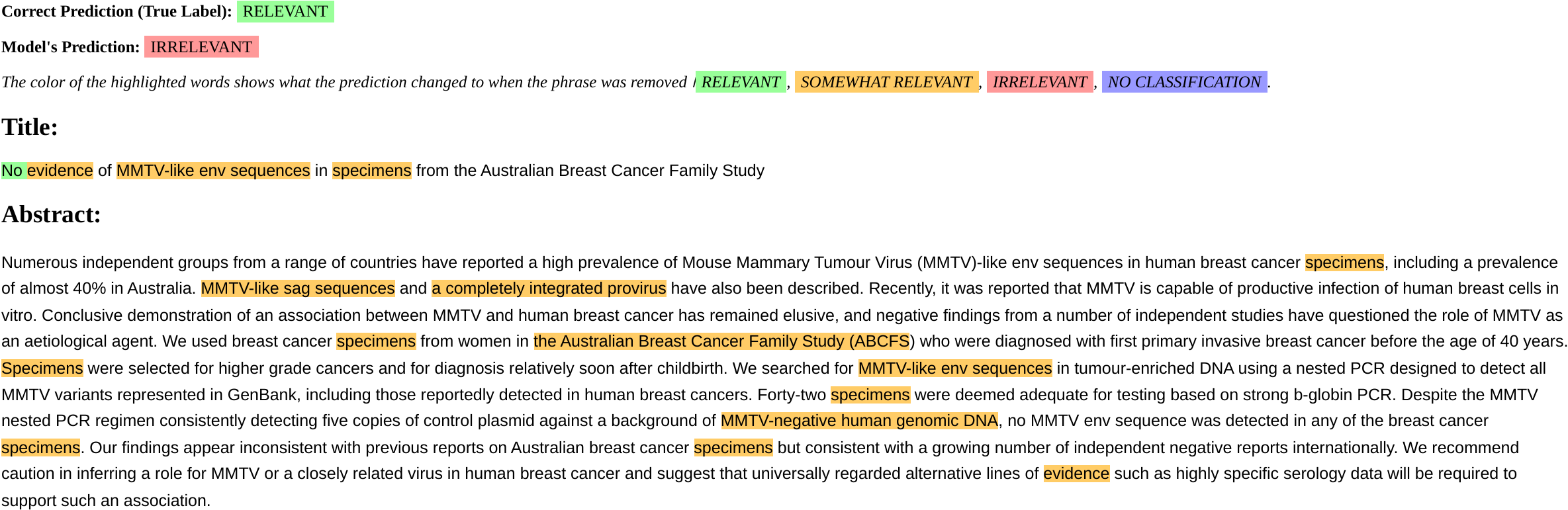}
    \caption{Word removal analysis for Paper 26 with Llama 3. The highlighted words represent those removed during analysis, with each color denoting the new class assigned after removal (green for \textsc{RELEVANT}, orange for \textsc{SOMEWHAT RELEVANT}, red for \textsc{IRRELEVANT}, and purple for \textsc{NO CLASSIFICATION}). Removing the word “No” shifted the paper classification to relevant, suggesting a bias toward positive findings. The same shift occurred when removing both key topic terms (e.g., “MMTV-like env sequences”) and unrelated words (e.g., “specimens”, “evidence”), highlighting inconsistent sensitivity, also observed in Qwen2.5.}
    \label{fig:P26_llama}
    \Description[Word removal analysis revealing Llama 3 bias toward positive findings in Paper 26]{Word removal analysis for Paper 26 with Llama 3 showing a title and abstract with color-coded highlights. Each highlighted word or phrase indicates the classification change when removed: green for relevant, orange for somewhat relevant, red for irrelevant, and purple for no classification. Removing the word “No” from the title shifted classification toward relevant, suggesting bias toward positive findings. Removing both domain-relevant terms like “MMTV-like env sequences” and clearly unrelated words like “specimens” and “evidence” produced similar shifts toward higher relevance, mirroring the inconsistent behavior observed with Qwen2.5.}
\end{figure}

\begin{figure}[ht]
    \centering
    \includegraphics[width=1\linewidth]{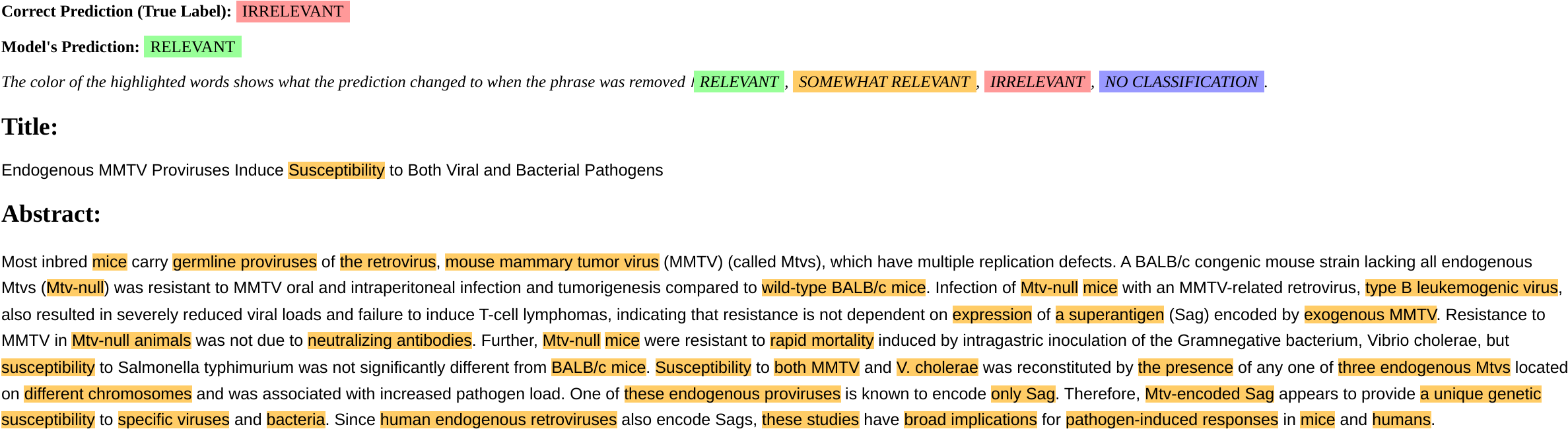}
    \caption{Word removal analysis for Paper 94 with Llama 3. The highlighted words represent those removed during analysis, with each color denoting the new class assigned after removal (green for \textsc{RELEVANT}, orange for \textsc{SOMEWHAT RELEVANT}, red for \textsc{IRRELEVANT}, and purple for \textsc{NO CLASSIFICATION}). Removing mouse-related terms (e.g., “mice”, “wild-type BALB/c mice”, “Mtv-null animals”) and even unrelated words (e.g., “susceptibility”, “V. cholerae”) reduced relevance scores, highlighting the model’s tendency to treat both key and random terms as important for classification.}
    \label{fig:P94_llama}
    \Description[Word removal analysis showing Llama 3 treating relevant and irrelevant terms similarly in Paper 94]{Word removal analysis for Paper 94 with Llama 3 showing a title and abstract with color-coded highlights. Each highlighted word or phrase indicates the classification change when removed: green for relevant, orange for somewhat relevant, red for irrelevant, and purple for no classification. Removing biologically meaningful terms such as “mice”, “wild-type BALB/c mice”, and “Mtv-null animals” decreased relevance scores as expected, but removing unrelated terms like “susceptibility” and “V. cholerae” produced comparable effects, demonstrating that the model weighs scientifically important and irrelevant terms similarly.}
\end{figure}

Figure~\ref{fig:P26_llama} shows Llama 3’s analysis of Paper 26. Removing the term “No” shifted the classification toward relevant, suggesting a potential bias toward positive findings. More concerningly, removing both domain-relevant terms (e.g., “MMTV-like env sequences”) and clearly unrelated words (e.g., “specimens,” “evidence”) produced similar shifts toward higher relevance—mirroring the behavior observed with Qwen2.5. A comparable pattern appears in Figure~\ref{fig:P94_llama}. Removing biologically meaningful terms such as “mice,” “wild-type BALB/c mice,” and “Mtv-null animals” decreased relevance scores, as expected. However, removing unrelated terms like “susceptibility” and “V. cholerae” produced comparable effects, indicating that the model weighs scientifically important and irrelevant terms similarly in its decision-making process. These results confirm that both Qwen2.5 and Llama 3 exhibit the same fundamental tendency: effective reasoning paired with inconsistent semantic reasoning. Consequently, while such models can serve as efficient tools for first-pass filtering in large-scale literature screening, they remain unsuitable for subsequent autonomous final decision-making tasks that require robust and interpretable scientific reasoning.

\section{Precision–Sensitivity Analysis Across All Classes}

\begin{figure}[H]
    \centering
    \includegraphics[width=0.95\linewidth]{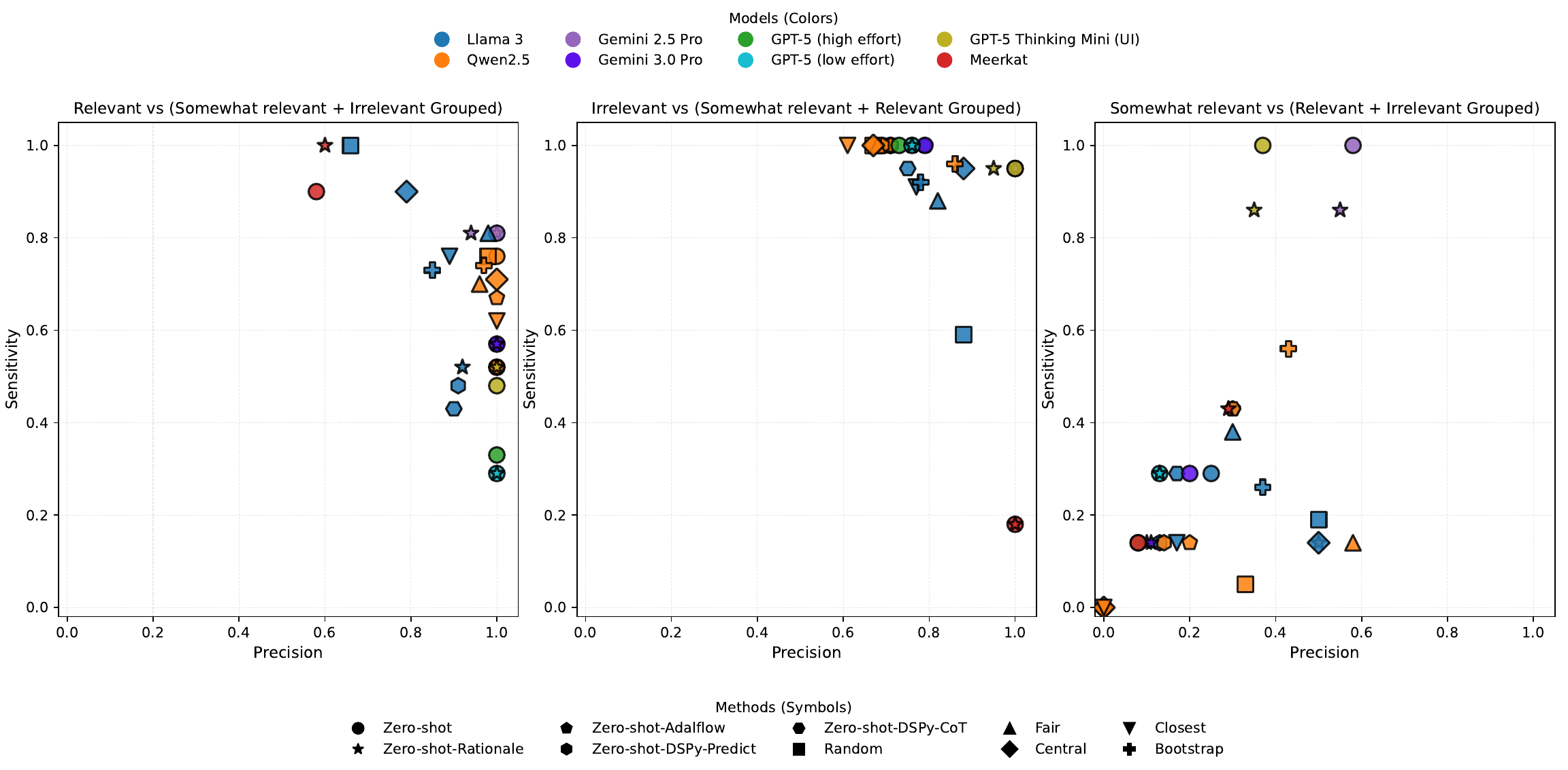}
    \caption{Binary performance for different groupings of the three classes. {\bf Left:} RELEVANT vs. SOMEWHAT RELEVANT grouped with IRRELEVANT. {\bf Middle:} IRRELEVANT vs. RELEVANT grouped with SOMEWHAT RELEVANT. {\bf Right:} SOMEWHAT RELEVANT vs. RELEVANT grouped with IRRELEVANT. Across all configurations, RELEVANT and IRRELEVANT classes are comparatively easy to distinguish, whereas SOMEWHAT RELEVANT remains the most challenging and discriminative class; notably, Qwen2.5 with BFS-RS (orange cross, right panel) stands out as the strongest SLM for this difficult class, explaining its superior overall F1 performance.}
    \label{fig:precisionSensTradeOff_All}
    \Description[Precision-sensitivity tradeoffs across three class groupings]{Three scatter plots compare precision and sensitivity when each of the three classes is treated as the positive class in turn. When RELEVANT is positive (left), most models achieve near-perfect precision with moderate sensitivity, indicating conservative detection behavior. When IRRELEVANT is positive (middle), nearly all models cluster near perfect precision and sensitivity, showing that this class is easiest to separate. When SOMEWHAT RELEVANT is positive (right), performance drops substantially across most models, revealing this class to be the most difficult and ambiguous; only a few configurations achieve moderate sensitivity with limited precision, while Qwen2.5 with BFS-RS performs strongest among smaller models.}
\end{figure}

Figure~\ref{fig:precisionSensTradeOff_All} presents the precision-sensitivity tradeoff plots for all classes, where each class (RELEVANT, SOMEWHAT RELEVANT, and IRRELEVANT) is treated as the positive class in turn.

When the RELEVANT class is treated as positive (left panel), this plot has already been discussed in Section~\ref{subsection:prec-sens}. In summary, most model configurations lie in the region of perfect precision with moderate sensitivity. This indicates that when these models classify a paper as RELEVANT, they are usually correct; however, the lower sensitivity reflects a conservative behavior, meaning that not all RELEVANT papers are captured. A clear tradeoff is observed for Llama 3, where different configurations prioritize either precision or sensitivity. When the IRRELEVANT class is treated as positive (middle panel), all models perform exceptionally well. This is expected, as IRRELEVANT papers are easier to distinguish from RELEVANT and SOMEWHAT RELEVANT papers, given that they largely consist of unrelated medical content outside the Microbe-Cancer Pair (HMTV/MMTV-like virus-Breast cancer) domain.

Finally, when the SOMEWHAT RELEVANT class is treated as positive (right panel), we observe important insights. With the exception of Gemini 2.5 Pro and GPT-5 Mini (including their Rationale variants), all models struggle significantly on this class. Although Gemini 2.5 Pro and GPT-5 Mini achieve high sensitivity, their precision remains modest, with maximum precision around 0.6 for Gemini 2.5 Pro. This highlights the inherent difficulty of the SOMEWHAT RELEVANT class and underscores the nuanced complexity of our classification task. Notably, Qwen2.5 with BFS-RS performs exceptionally well on this class compared to other models and is the best SLM configuration for it. This explains the high overall three-class performance of Qwen2.5 with BFS-RS, as noted in Section~\ref{subsec:three-class}, where it achieved the highest F1 score among the SLM configurations.

Overall, these precision-sensitivity tradeoff plots highlight the varying challenges across the three classes. While RELEVANT and IRRELEVANT papers are generally easier for models to classify, the SOMEWHAT RELEVANT class remains a difficult and nuanced category, reflecting the subtle and nuanced nature of our classification criteria. The strong performance of Qwen2.5 with BFS-RS on this class demonstrates the impact of programmatic prompt optimization in handling subtle distinctions, ultimately explaining its superior overall F1 score. These results underscore the importance of carefully evaluating each class individually to understand model behavior and guide future improvements.

\end{document}